\documentclass[twocolumn,prl,superscriptaddress,longbibliography]{revtex4-1}
\usepackage{graphicx,amsmath,amssymb,braket,mathtools}
\usepackage[colorlinks=true, citecolor=blue, urlcolor=blue, linkcolor=red]{hyperref}
\renewcommand{\section}[1]{{\par\it #1.---}\ignorespaces}

\begin{document}
\title{Quantum Mpemba Effect Induced by Non-Markovian Exceptional Points}
\author{Ze-Zhou Zhang}
\affiliation{Key Laboratory of Quantum Theory and Applications of Ministry of Education, Lanzhou Center for Theoretical Physics and Key Laboratory of Theoretical Physics of Gansu Province, and School of Physical Science and Technology, Lanzhou University, Lanzhou 730000, China}
\author{Hong-Gang Luo}
\affiliation{Key Laboratory of Quantum Theory and Applications of Ministry of Education, Lanzhou Center for Theoretical Physics and Key Laboratory of Theoretical Physics of Gansu Province, and School of Physical Science and Technology, Lanzhou University, Lanzhou 730000, China}
\author{Wei Wu}
\email{wuw@lzu.edu.cn}
\affiliation{Key Laboratory of Quantum Theory and Applications of Ministry of Education, Lanzhou Center for Theoretical Physics and Key Laboratory of Theoretical Physics of Gansu Province, and School of Physical Science and Technology, Lanzhou University, Lanzhou 730000, China}

\begin{abstract}
Quantum Mpemba effect describes an anomalous phenomenon of accelerated relaxation, which is of fundamental interest in the field of nonequilibrium thermodynamics. Conventional theories on this phenomenon strongly rely on the Born-Markovian approximation resulting in a Lindblad-type master equation whose evolution is governed by a Liouvillian superoperator. It has been demonstrated that exceptional points of the Liouvillian superoperator can induce the Mpemba effect in Markovian regimes. Moving beyond this Markovian limit, we here propose a mechanism for observing the quantum Mpemba effect in a general non-Markovian relaxation process by means of non-Markovian exceptional points. We verify the feasibility of this mechanism within a dissipative quantum harmonic oscillator model, which is exactly solvable and experimentally practical. Providing new insight into the interesting nonequilibrium dynamics, our work paves a way to accelerate the transfer of energy and information in quantum systems.
\end{abstract}
\maketitle

\section{Introduction}\label{sec:sec1}
The Mpemba effect describes a counterintuitive phenomenon: a hot system freezes faster than a cold one, when both are placed in the same environment~\cite{EBMpemba_1969,10.1119/1.1975687,10.1119/1.3490015}. In recent years, the concept of the Mpemba effect has been generalized to microscopic quantum systems, giving rise to the quantum Mpemba effect (QMPE)~\cite{PhysRevLett.127.060401,PhysRevLett.131.080402,PhysRevLett.134.220403,PhysRevLett.133.010401,PhysRevLett.133.010402,PhysRevLett.134.220402,Zhang2025,PhysRevA.110.022213,Longhi2025mpembaeffectsuper,10.1063/5.0234457,PhysRevResearch.6.033330}. The QMPE refers to an anomalous phenomenon where a far-from-equilibrium state relaxes to equilibrium more quickly than a state closer to equilibrium. Theoretically, the QMPE opens an alternative avenue to understand thermalization. From an application perspective, an accelerated relaxation may improve the discharge efficiency of quantum battery~\cite{PhysRevLett.134.220402} and optimize quantum heat engines~\cite{PhysRevA.110.042218}. Thus, much effort has been devoted to identifying and understanding the underlying mechanism of QMPE.

While numerous explanations have been proposed~\cite{doi:10.1073/pnas.1701264114,PhysRevLett.133.136302,PhysRevLett.125.110602,Bechhoefer2021,liu2025generalstrategyrealizingmpemba,PhysRevResearch.5.043036,summer2025resourcetheoreticalunificationmpemba,LIU20253991,Yu2025,PhysRevLett.133.140405}, the QMPE still grabs much attraction. In previous studies, to explain the QMPE within the theory of open quantum system, the Born-Markovian approximation is employed~\cite{PhysRevLett.134.220402,doi:10.1073/pnas.1701264114,PhysRevLett.127.060401,g94p-7421,PhysRevResearch.5.043036,qj8n-k5j2,PhysRevLett.133.140404}. Under this approximation, the reduced dynamics can be expressed in a Lindblad master equation~\cite{Breuer,PhysRevA.98.042118,PhysRevA.100.062131}. This Lindbladian evolution shows that the relaxation timescale is determined by the slowest eigenmode of the Liouvillian operator~\cite{PhysRevLett.127.060401}. This mechanism reveals the occurrence of Liouvillian exceptional point (LEP), where two or more eigenenergies become degenerate and the corresponding eigenvectors coalesce, is able to induce the QMPE~\cite{PhysRevResearch.5.043036,Zhang2025}.

However, the above LEP-induced QMPE is largely confined to the Born-Markov approximation, which is only valid in the weak-coupling regime or when the bath energy spectrum is flat \cite{RevModPhys.88.021002,RevModPhys.89.015001,PhysRevE.90.022122,PhysRevLett.109.170402}. If the coupling becomes sufficiently strong, the non-Markovianity, which is commonly encoded in the memory kernel of a time-nonlocal equation of motion~\cite{Breuer}, invalidates the traditional Lindblad master equation as well as the concept of LEP~\cite{Lin2025}. This non-Markovianity is not merely a theoretical extension but a necessary reflection of experimental reality. Recent advances in the superconducting circuit platform have clearly demonstrated non-Markovian dynamics~\cite{jk6y-55xp}, making the non-Markovianity essential in real quantum devices. There have been a few studies on non-Markovian QMPE~\cite{PhysRevLett.134.220403,strachan2025acceleratedcalculationimpuritygreens,alyürük2025thermodynamiclimitsmpembaeffect,5xrr-x2rm}, these studies typically employed the time-convolutionless master equation with a time-dependent Liouvillian operator~\cite{PhysRevB.83.115416,PhysRevLett.104.070406}. First, the time-convolutionless master equation is not the only non-Markovian dynamics method~\cite{RevModPhys.89.015001}, studying the QMPE from other perspectives can provide completely different insights. Second, the time-convolutionless master equation complicates the application of traditional spectral analysis techniques for identifying LEPs in non-Markovian scenarios. A gap still remains in understanding the relationship between LEPs and the non-Markovian QMPE.

In this Letter, we aim to address this gap. To achieve this goal, one first needs to generalize the concept of LEP to the non-Markovian dynamics, which has been realized by applying the pseudomode master equation approach~\cite{PhysRevLett.120.030402,PhysRevLett.123.090402,PhysRevA.55.2290}. This approach incorporates the non-Markovian effect through introducing auxiliary pseudomodes. A similar idea is used in the hierarchical equations of motion method where the non-Markovianity is integrated into auxiliary density operators~\cite{doi:10.1143/JPSJ.58.101,10.1063/5.0011599,PhysRevA.98.012110,PhysRevA.98.032116}. By doing so, the original non-Markovian dynamics becomes dynamically equivalent to a system-pseudomode Lindbladian evolution. Using this pseudomode master equation approach, we uncover a notable QMPE induced by non-Markovian LEPs, which are inaccessible under the Born-Markovian approximation. Our study establishes a unified framework that bridges the connection between the LEPs and the QMPE within both Markovian and non-Markovian dynamics.

\section{Non-Markovian dynamics}\label{sec:sec2}
Let us consider a general open quantum system, which is coupled to a dissipative bosonic bath via a linear interaction as ($\hbar=1$)
\begin{equation}\label{S1}
    \hat{H}_{\text{tot}}=\hat{H}_{\text{s}}+\hat{H}_{\text{b}}+\hat{S}\hat{B}^{\dagger}+\hat{S}^{\dagger}\hat{B},
\end{equation}
where $\hat{H}_{\text{s}}$ is the system, and $\hat{S}$ denote the dissipation operator. The bath Hamiltonian is $\hat{H}_{\text{b}}=\sum_{k}\omega_{k}\hat{b}_{k}^{\dagger}\hat{b}_{k}$ and $\hat{B}=\sum_{k}g_{k}\hat{b}_{k}$, where $\hat{b}_{k}$ and $\hat{b}_{k}^{\dagger}$ are the bosonic annihilation and creation operators of the $k$th bosonic mode with the corresponding frequency $\omega_{k}$, respectively. Parameter $g_{k}$ quantifies the coupling strength. The frequency dependence of the coupling strengths is encoded into the spectral density as $J(\omega)\equiv\sum_{k}|g_{k}|^{2}\delta(\omega-\omega_{k})$. The explicit expressions of $\hat{H}_{\text{s}}$, $\hat{S}$, and $J(\omega)$ are not addressed here, because our formula is universal to their forms.

\begin{figure}\label{Figure1}
\centering
\includegraphics[angle=0,width=0.45\textwidth]{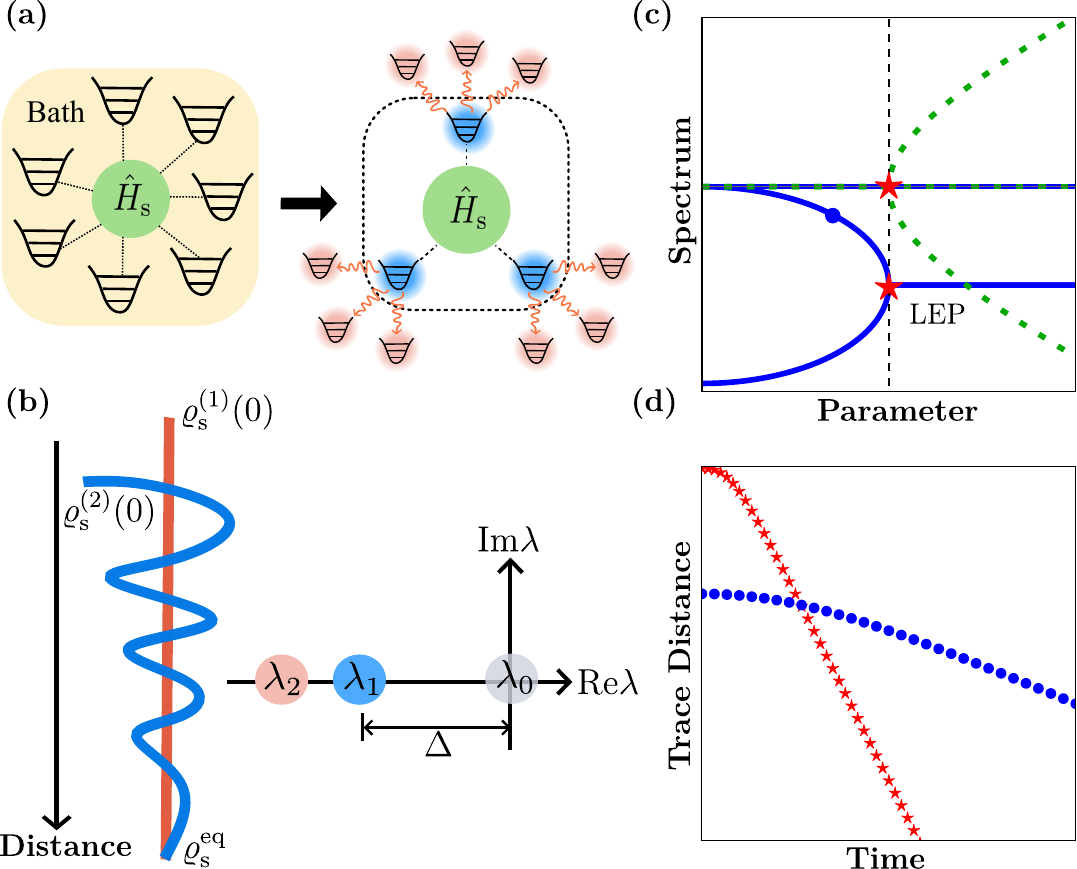}
\caption{(a) Sketch of the pseudomode master equation approach. (b) The cartoon of QMPE: an initial far-from-equilibrium state $\varrho_{\text{s}}^{(1)}(0)$ can relax toward the equilibrium state faster than a closer-to-equilibrium initial state $\varrho_{\text{s}}^{(2)}(0)$. (c) The Liouvillian spectral analysis, the blue solid lines are $\text{Re}(\lambda_{\ell})$ and the green dashed lines are $\text{Im}(\lambda_{\ell})$. The LEPs are marked by red stars and one non-LEP is represented by the blue circle. (d) A QMPE can be observed by using non-Markovian LEPs.}\label{fig:fig1}
\end{figure}

Assuming the initial state of the whole Hamiltonian is $\rho_{\text{tot}}(0)=\rho_{\text{s}}(0)\otimes\rho_{\text{b}}(0)$ with $\rho_{\text{b}}(0)=\bigotimes_{k}|0_{k}\rangle\langle 0_{k}|$ being the vacuum state of the bath, the reduced dynamics of system is determined by
\begin{equation}\label{eq:eq2}
\rho_{\text{s}}(t)=\text{Tr}_{\text{b}}[e^{-it\hat{H}_{\text{tot}}}\rho_{\text{s}}(0)\otimes\rho_{\text{b}}(0)e^{it\hat{H}_{\text{tot}}}].
\end{equation}
One finds that $\rho_{\text{s}}(t)$ is completely determined by the two-time bath correlation function $C(t)=\int_{0}^{\infty}d\omega J(\omega)e^{-i\omega t}$ due to the Gaussianity of the bosonic bath~\cite{SupplementalMaterial}, which guarantees the reduced dynamics is tractable~\cite{PhysRevLett.113.200403}. Even though only the two-time correlation function is involved, due to the time-nonlocal convolution terms, which collect the non-Markovian effect, $\rho_{\text{s}}(t)$ still lacks a very simple expression. Fortunately, if this bath correlation function can be (at least approximately) written as a finite sum of exponentials
\begin{equation}\label{eq:eq3}
C(t)=\sum_{i=1}^{N}\alpha_{i}^{2}e^{-i\Omega_{i}t-\frac{1}{2}\gamma_{i}t},
\end{equation}
one proves that $\rho_{\text{s}}(t)$ is dynamically equivalent to $\varrho_{\text{s}}(t)=\text{Tr}_{\text{p}}[\rho_{\text{sp}}(t)]$. Here, $\rho_{\text{sp}}(t)$ is governed by~\cite{Lin2025,PhysRevLett.120.030402,Lambert2019}
\begin{equation}\label{eq:eq4}
\dot{\rho}_{\text{sp}}(t)=\mathcal{\hat{L}}\rho_{\text{sp}}(t)=-i[\hat{H}_{\text{sp}},\rho_{\text{sp}}(t)]+\sum_{i=1}^{N}\gamma_{i}\mathcal{\hat{D}}_{\hat{a}_{i}}[\rho_{\text{sp}}(t)],
\end{equation}
where $\mathcal{\hat{L}}$ is an extended Liouvillian superoperator and
\begin{equation}
\hat{H}_{\text{sp}}= \hat{H}_{\text{s}}+\sum_{i=1}^{N}\Omega_{i}\hat{a}^{\dagger}_{i}\hat{a}_{i}+\sum_{i=1}^{N}(\alpha_{i}\hat{S}\hat{a}^{\dagger}_{i}+\text{H}.\text{c}.),
\end{equation}
which is an enlarged system consisting of the original system $\hat{H}_{\text{s}}$ and the pseudomodes. Operator $\hat{a}_{i}$ is the annihilation operator of the $i$th pseudomode with frequency $\Omega_{i}$. The Lindblad dissipator is defined as $\mathcal{\hat{D}}_{\hat{a}_{i}}[\bullet]=\hat{a}_{i}\bullet \hat{a}^{\dagger}_{i}-\frac{1}{2}\{\hat{a}^{\dagger}_{i}\hat{a}_{i},\bullet\}$. The initial-state condition for Eq.~(\ref{eq:eq4}) is $\rho_{\text{sp}}(0)=\rho_{\text{s}}(0)\otimes\rho_{\text{p}}(0)$ with $\rho_{\text{p}}(0)=\bigotimes_{i}|0_{i}\rangle\langle 0_{i}|$.

This pseudomode master equation approach is also known as the Markovian embedding technique~\cite{RevModPhys.89.015001,PhysRevA.55.2290,PhysRevE.81.011136,10.1063/1.3532408,PhysRevA.90.032114}, which represents a non-Markovian evolution as a projection of a higher dimensional Markovian dynamics. Non-Markovian effects can be captured by the auxiliary degrees of freedom: when the energy
or the information flows from the system to the pseudomodes, it can be partially transferred back to the system via the system-pseudomodes interaction. Numerical or analytical treatments of Eq.~(\ref{eq:eq4}) are typically more efficient than that of Eq.~(\ref{eq:eq2}), thus this method is a powerful tool to investigate the non-Markovian dynamics and has been widely used in many previous studies~\cite{c91x-bhqw,Sun2025,Zeng2025,PhysRevB.110.195148}. More importantly, this method allows us to perform conventional spectral analysis for the enlarged Liouvillian superoperator $\mathcal{\hat{L}}$~\cite{Lin2025}. Compared with the traditional Markovian LEPs, the introduction of auxiliary pseudomodes, can result in additional LEPs, which greatly modify the nonequilibrium dynamics of relaxation. Thus, the LEPs of $\mathcal{\hat{L}}$ are called non-Markovian LEPs, which are fundamentally different from the Markovian LEPs. We want to emphasize Eq.~(\ref{eq:eq3}) is not a very stringent restriction and is widely needed in Refs.~\cite{doi:10.1143/JPSJ.75.082001,PhysRevE.75.031107,PhysRevLett.113.150403}. Using certain numerical fitting techniques~\cite{dattani2012optimalrepresentationbathresponse,10.1063/1.5100102,10.1063/1.4893931}, $C(t)$ can be conveniently expressed as Eq.~(\ref{eq:eq3}). The limitation of our employed method mainly comes from the numerical cost of solving Eq.~(\ref{eq:eq4}) when $N$ is very large. And this method is commonly used to investigate Gaussian open systems, it is still an unknown question whether this method can be feasible in the non-Gaussian bath models, in which multitime bath correlation functions should be separately taken into account~\cite{PhysRevA.98.032116,10.1063/1.5018725}.

\section{Non-Markovian LEPs and QMPE}
If $\mathcal{\hat{L}}$ is diagonalizable, its complex eigenvalues $\lambda_{\ell}$, the corresponding right and left eigenmatrices, $r_{\ell}$ and $l_{\ell}$, satisfy  $\mathcal{\hat{L}}r_{\ell}=\lambda_{\ell}r_{\ell}$ and $\mathcal{\hat{L}}^{\dagger}l_{\ell}=\lambda_{\ell}^{*}l_{\ell}$, respectively. As long as the eigenvalues and the eigenmatrices are known, $\rho_{\text{sp}}(t)$ is expressed as
\begin{equation}
\rho_{\text{sp}}(t)=\sum_{\ell=0}^{M^{2}-1}\text{Tr}[l_{\ell}\rho_{\text{sp}}(0)]e^{\lambda_{\ell}t}r_{\ell},
\end{equation}
where $l_{0}$ is an identity, $\lambda_{0}=0$, $r_{0}=\rho^{\text{eq}}_{\text{sp}}$ with $\mathcal{\hat{L}}\rho^{\text{eq}}_{\text{sp}}=0$ is the steady-state solution of Eq~(\ref{eq:eq4}), and $M\equiv\text{Dim}(\hat{H}_{\text{sp}})$ is the dimension of the enlarged system. $\text{Re}(\lambda_{\ell})$ are known to be negative and represent the relaxation rates toward the long-time steady state. Arranging them in an ascending order $|\text{Re}(\lambda_{1})|\leq |\text{Re}(\lambda_{2})|\dots\leq|\text{Re}(\lambda_{M^{2}-1})|$, one sees the relaxation dynamics of $\rho_{\text{sp}}(t)$ in the long-time regime is mainly governed by $\rho_{\text{sp}}(t)\propto e^{-\Delta t}r_{1}$, where $\Delta\equiv-\text{Re}(\lambda_{1})$ is the Liouvillian spectral gap.

Using the above result, in the long-time regime, one immediately concludes that the trace distance between $\varrho_{\text{s}}(t)$ and $\varrho_{\text{s}}^{\text{eq}}=\text{Tr}_{\text{p}}(\rho^{\text{eq}}_{\text{sp}})$ is determined by (see Supplemental Material~\cite{SupplementalMaterial} for the proof)
\begin{equation}\label{eq:eq7}
||\varrho_{\text{s}}(t)-\varrho_{\text{s}}^{\text{eq}}||_{\text{TD}}\propto e^{-\Delta t},
\end{equation}
where $||A||_{\text{TD}}\equiv\frac{1}{2}\text{Tr}\sqrt{AA^{\dagger}}$~\cite{Nielsen} is the trace norm. Equation~(\ref{eq:eq7}) means the Liouvillian spectral gap sets the timescale for the slowest relaxation to the steady state as $\tau_{\text{R}}\sim 1/\Delta$. Via increasing the value of $\Delta$, the relaxation timescale can be accordingly compressed. When the slowest decay mode $\{\lambda_{1},r_{1}\}$ coalesces with a faster decay mode $\{\lambda_{2},r_{2}\}$ at the LEP, the Liouvillian spectral gap is maximized, which speeds up the relaxation toward the steady state. Using these discussions, a scheme for observing the QMPE can be established: choosing two different initial states $\varrho_{\text{s}}^{(1)}(0)$ and $\varrho_{\text{s}}^{(2)}(0)$ with $\varrho_{\text{s}}^{(2)}(0)$ being closer to the long-time steady state $\varrho_{\text{s}}^{\text{eq}}$, then $\varrho_{\text{s}}^{(1)}(0)$ and $\varrho_{\text{s}}^{(2)}(0)$ are designed to evolve under the guidances of $\mathcal{\hat{L}}$ with and without an LEP, respectively. As plotted in Fig.~\ref{fig:fig1}, although the distance $||\varrho_{\text{s}}^{(1)}(0)-\varrho_{\text{s}}^{\text{eq}}||_{\text{TD}}$ is larger, $\varrho_{\text{s}}^{(1)}(t)$ would relax at a faster relaxation rate induced by the occurrence of the LEP. Via observing the presence (or absence) of an intersection of the distances, one can determine the occurrence of QMPE. Such a QMPE is generated by the spectral singularities of $\mathcal{\hat{L}}$, which can not be predicted by the Born-Markovian approach.

\begin{figure}\label{Figure2}
\centering
\includegraphics[angle=0,width=0.475\textwidth]{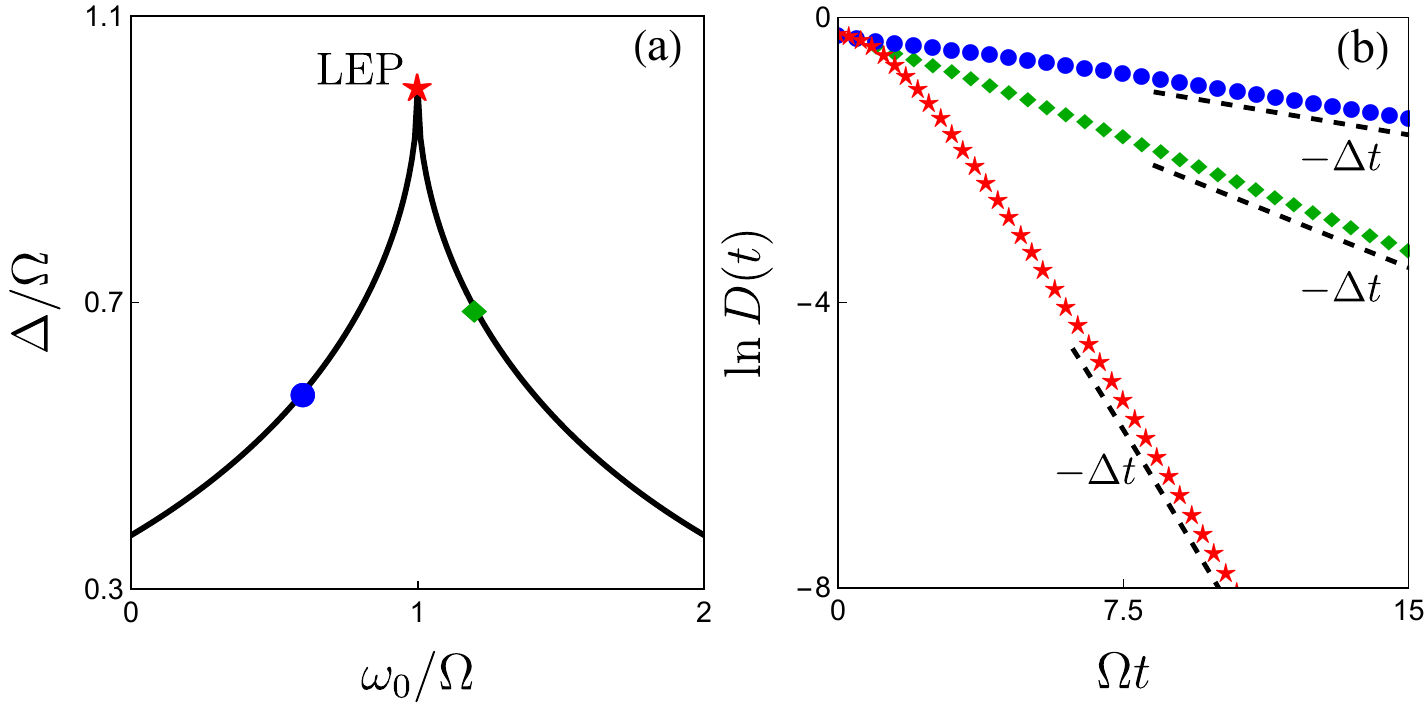}
\caption{(a) The Liouvillian spectral gap is plotted as a function of $\omega_{0}$ with fixed $\gamma=4\alpha$. It is revealed that $\Delta$ reaches its maximum value if the condition of a LEP occurring, i.e.,  $\omega_{0}=\Omega$, is satisfied. The corresponding trace distance $\ln D(t)$ is plotted as a function of $\Omega t$ with different frequencies: $\omega_{0}=0.6\Omega$ (blue rectangles), $\omega_{0}=\Omega$ (red stars) and $\omega_{0}=1.2\Omega$ (green diamonds). The dashed lines are analytical results from the Liouvillian spectral gap with $\Delta=\frac{1}{4}\gamma-\text{Re}(\kappa)$. The explicit expression of $D(t)$ is given by Eq.~(\ref{eq:eq12}) and the initial state is a coherent state with $\xi=1$. Other parameters are chosen as $\Omega=1~\text{cm}^{-1}$ and $\alpha/\Omega=1$.}\label{fig:fig2}
\end{figure}

To realize the above scheme, one needs to find the eigenvalues and the corresponding eigenmatrices, whose analytical expressions are generally unknown. However, for a dissipative quantum harmonic oscillator, i.e., $\hat{H}_{\text{s}}=\omega_{0}\hat{a}^{\dagger}_{0}\hat{a}_{0}$ and $\hat{S}=\hat{a}_{0}$, the analytical expressions for $\lambda_{\ell}$ and $\rho_{\text{sp}}(t)$ can be derived by using the Bogoliubov transformation and the su(1,1) Lie algebra~\cite{2lgr-34qp}. This model is exactly solvable~\cite{PhysRevA.82.012105}, it can be employed as the benchmark to check the correctness of our proposed scheme. We find~\cite{SupplementalMaterial}
\begin{equation}\label{eq:eq8}
\lambda_{\pmb{mn}}=\sum_{\imath=0}^{N}[i\text{Im}(\tilde{\lambda}_{\imath})(m_{\imath}-n_{\imath})+\text{Re}(\tilde{\lambda}_{\imath})(m_{\imath}+n_{\imath})],
\end{equation}
where $\pmb{mn}\equiv\{m_{0},n_{0},m_{1},n_{1},...,m_{N},n_{N}\}$ is an index with $m_{\imath},n_{\imath}\in \mathbb{N}$, and $\tilde{\lambda}_{\imath}$ are the roots of the following characteristic polynomial
\begin{equation}\label{eq:eq9}
\begin{split}
Q(\lambda)&=(i\omega_{0}+\lambda)\prod_{i=1}^{N}\left(i\Omega_{i}+\frac{\gamma_{i}}{2}+\lambda\right)\\
&+\sum_{i=1}^{N}\alpha_{i}^{2}\prod_{i'=1,i'\neq i}^{N}\left(i\Omega_{i'}+\frac{\gamma_{i'}}{2}+\lambda\right).
\end{split}
\end{equation}
The formal expression for $\rho_{\text{sp}}(t)$ is given in Supplemental Material~\cite{SupplementalMaterial}.

By choosing a suitable set of $\{\omega_{0},\Omega_{i},\gamma_{i},\alpha_{i}\}$, LEPs can be found. One can fix the frequency of the system, but adjust the parameters of the bath correlation function, or conversely, fix $\{\alpha_{i},\gamma_{i},\Omega_{i}\}$, but change $\omega_{0}$. The first strategy is adopted in Ref.~\cite{PhysRevResearch.5.043036}. However, the second one is closer to the original definition of the Mpemba effect, in which the ``hotter" and ``colder" systems are placed in the same environment. Taking $N=1$ as an example, we find the two roots of Eq.~(\ref{eq:eq9}) are $\tilde{\lambda}_{0,1}=-\frac{1}{4}\gamma\pm\kappa-\frac{i}{2}(\omega_{0}+\Omega)$ with $\kappa=\sqrt{[\gamma+2i(\Omega-\omega_{0})]^{2}/16-\alpha^{2}}$. This result means the first three eigenvalues of $\lambda_{m_{0}n_{0}m_{1}n_{1}}$ are $\lambda_{0}=\lambda_{0000}=0$, $\lambda_{1}=\lambda_{1000}=-\frac{1}{4}\gamma+\kappa-\frac{i}{2}(\omega_{0}+\Omega)$ and $\lambda_{2}=\lambda_{0010}=-\frac{1}{4}\gamma-\kappa-\frac{i}{2}(\omega_{0}+\Omega)$. The corresponding Liouvillian spectral gap is $\Delta=\frac{1}{4}\gamma-\text{Re}(\kappa)$. Clearly, a LEP occurs at $\gamma=4\alpha$ and $\omega_{0}=\Omega$, where $\lambda_{1}=\lambda_{2}=-\frac{1}{4}\gamma-i\omega_{0}$. As displayed in Fig.~\ref{fig:fig2}, when the LEP occurs, $\Delta$ indeed reaches its maximum value.

\section{Comparison with Born-Markovian results}
For the sake of completeness, we provide a proof that the pseudomode master equation of Eq.~(\ref{eq:eq4}) reduces to the standard Lindblad master equation under the Born-Markovian approximation. As discussed in Refs.~\cite{PhysRevA.98.032116,PhysRevLett.105.240403}, $\gamma_{i}$ are related to the memory time of bath. If $\gamma_{i}\to \infty$, $C(t)$ vanishes and the dynamics becomes Markovian. Applying both the limit of $\gamma_{i}\to \infty$ and the Born approximation $\rho_{\text{sp}}(t)\approx \varrho_{\text{s}}(t)\otimes \rho_{\text{p}}(0)$, we prove that Eq.~(\ref{eq:eq4}) naturally degrades to~\cite{SupplementalMaterial}
\begin{equation}\label{eq:eq10}
\dot{\varrho}_{\text{s}}^{\text{M}}(t)=\mathcal{\hat{L}}_{\text{M}}\varrho_{\text{s}}^{\text{M}}(t)=-i[\hat{H}'_{\text{s}},\varrho_{\text{s}}^{\text{M}}(t)]+\gamma_{\text{M}}\mathcal{\hat{D}}_{\hat{a}_{0}}[\varrho^{\text{M}}_{\text{s}}(t)],
\end{equation}
where $\hat{H}'_{\text{s}}=\hat{H}_{\text{s}}+\hat{H}_{\text{LS}}$ with $\hat{H}_{\text{LS}}=\sum_{i}4\alpha_{i}^{2}\Omega_{i}/\gamma_{i}^{2}\hat{a}_{0}^{\dagger}\hat{a}_{0}$ being the Lamb shift term, and $\gamma_{\text{M}}=\sum_{i}4\alpha_{i}^{2}/\gamma_{i}$ is the Markovian decay rate. The corresponding eigenvalues of $\mathcal{\hat{L}}_{\text{M}}$ are $\lambda_{\pmb{mn}}^{\text{M}}=-\frac{1}{2}\gamma_{\text{M}}(m_{0}+n_{0})-i\omega_{0}(m_{0}-n_{0})$ with $m_{0},n_{0}\in \mathbb{N}$. Equation~(\ref{eq:eq10}) has a standard Lindblad solution whose analytical expression for $\varrho_{\text{s}}^{\text{M}}(t)$ is given in Supplemental Material~\cite{SupplementalMaterial}. The above result implies there is no LEP and the spectral gap $\Delta_{\text{M}}=\frac{1}{2}\gamma_{\text{M}}$ remains constant in the Markovian case.

Our proposed scheme offers two key advantages over the traditional method using Markovian LEPs to observe the QMPE. First, it enables the realization of an LEP-induced QMPE in regimes where the Born-Markovian theory fails. As demonstrated in Supplemental Material~\cite{SupplementalMaterial}, the non-Markovian-LEP-induced QMPE can be used to accelerate the discharging of a quantum battery. This is an advantage that vanishes under the  Born-Markovian approximation. Second, also shown in Supplemental Material~\cite{SupplementalMaterial}, thanks to the additional tunability of pseudomodes, our condition for forming LEPs is substantially relaxed compared with the Markovian LEP condition for qubit systems, making it experimentally feasible using trapped-ion platforms~\cite{Zhang2025}.

\section{Exemplification}
\begin{figure}
\includegraphics[angle=0,width=0.475\textwidth]{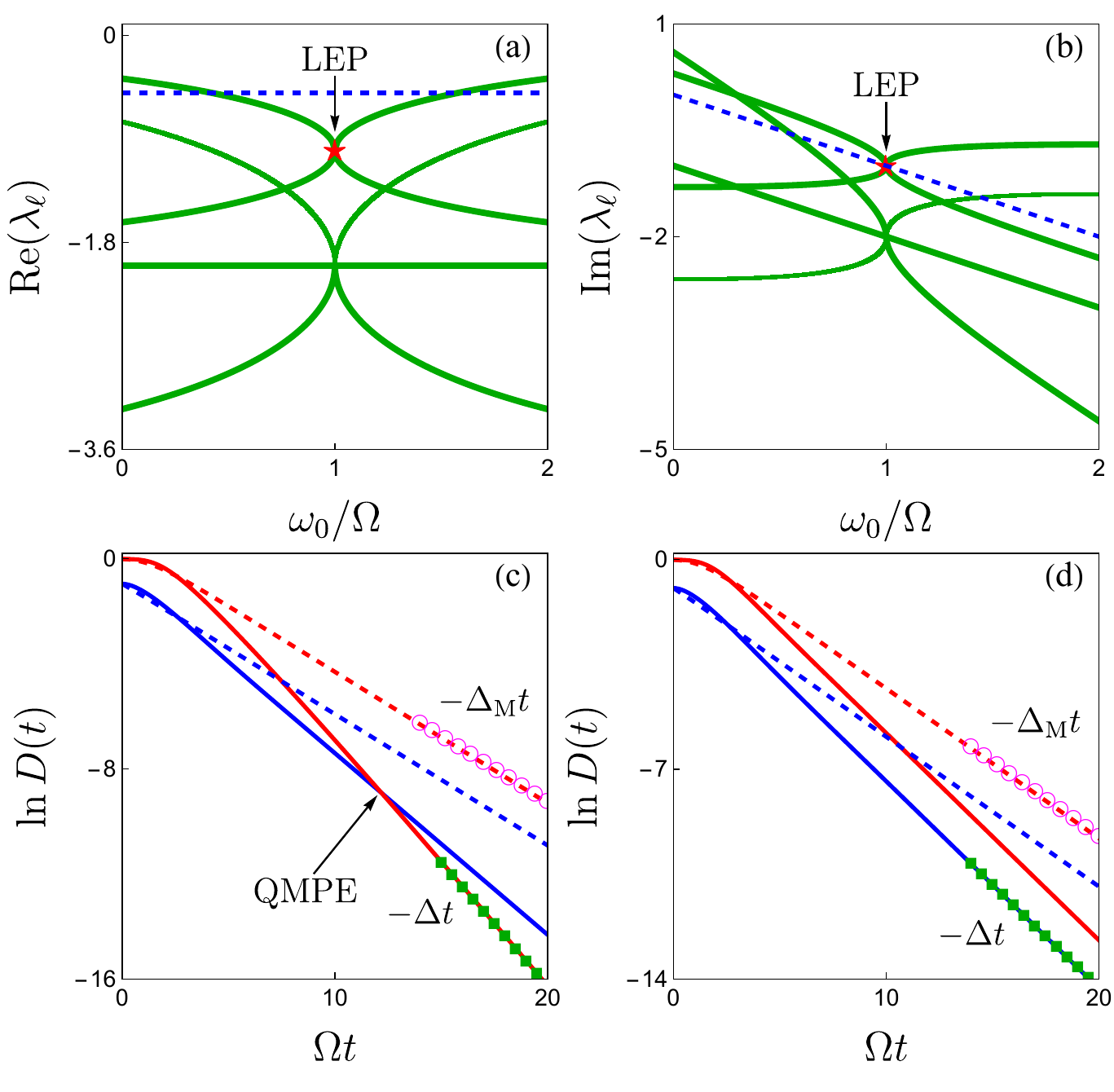}
\caption{Real (a) and imaginary (b) parts of the first five nonzero eigenvalues versus $\omega_{0}$ with fixed $\gamma=4\alpha$. The red star corresponds to the LEP where $\lambda_{1}=\lambda_{2}$. The green solid lines are non-Markovian results from $\mathcal{\hat{L}}$ and blue dashed lines are the Markovian result from $\mathcal{\hat{L}}_{\text{M}}$. (c) The corresponding trace distance $\ln D(t)$ versus $\Omega t$ with different initial-state parameters and frequencies: $\xi=0.4$ and $\omega_{0}=1.2\Omega$ (blue lines), $\xi=2$ and $\omega_{0}=\Omega$ (red lines). The solid lines are non-Markovian results, while the dashed lines are Markovian results. The green squares (magenta circles) are analytical results from the Liouvillian spectral gap with $\Delta=\frac{1}{4}\gamma-\text{Re}(\kappa)$ ($\Delta_{\text{M}}=\frac{1}{2}\gamma_{\text{M}}$). Subfigure (d) is the same as (c), but $\xi=0.4$ and $\omega_{0}=0.8\Omega$ (blue lines), $\xi=2$ and $\omega_{0}=1.2\Omega$ (red lines). Other parameters are chosen as $\alpha=\Omega$ and $\Omega=1~\text{cm}^{-1}$.}\label{fig:fig3}
\end{figure}
To verify the feasibility of our non-Markovian-LEP-induced QMPE scheme, we consider the dissipative quantum harmonic oscillator model. We assume the system is initially prepared in a coherent state $\rho_{\text{s}}(0)=|\xi\rangle\langle\xi|$ with $\xi\in \mathbb{R}$, and the spectral density is $J(\omega)=\frac{1}{2\pi}\frac{\Gamma \Lambda^{2}}{(\omega-\Omega)^{2}+\Lambda^{2}}$, where $\Gamma$ is the coupling strength and $\Lambda$ denotes the spectral width. For such a Lorentzian spectral density, the bath correlation function reads $C(t)=\frac{1}{2}\Gamma\Lambda e^{-(\Lambda+i\Omega)t}$, which naturally satisfies Eq.~(\ref{eq:eq3}). Solving Eq.~(\ref{eq:eq4}) and tracing out the pseudomode degrees of freedom, we find~\cite{SupplementalMaterial}
\begin{equation}\label{eq:eq11}
\varrho_{\text{s}}(t)=|\xi P(t)\rangle\langle\xi P(t)|,
\end{equation}
where $P(t)=e^{-i(\omega_{0}+\Omega)t/2-\gamma t/4}\{\cosh(\kappa t)+[\gamma+2i(\Omega-\omega_{0})]\sinh(\kappa t)/(4\kappa)\}$. This result is exactly the same with the rigorous result from the coherent-state path-integral method~\cite{PhysRevA.82.012105}. It is well known that an exact dynamics of an open quantum system is essentially non-Markovian~\cite{Rivas_2014,Breuer}. To clarify this point, the Bures distance is used to demonstrate the non-Markovianity~\cite{PhysRevLett.103.210401,PhysRevA.102.022228} of the dynamics described by Eq.~(\ref{eq:eq11})~\cite{SupplementalMaterial}. The trace distance between $\varrho_{\text{s}}(t)$ and $\varrho_{\text{s}}^{\text{eq}}$ is
\begin{equation}\label{eq:eq12}
D(t)=||\varrho_{\text{s}}(t)-\varrho_{\text{s}}^{\text{eq}}||_{\text{TD}}=\sqrt{1-e^{-|\xi P(t)|^{2}}}.
\end{equation}

We choose two initial coherent states $\varrho_{\text{s}}^{(1,2)}(0)=|\xi_{1,2}\rangle\langle\xi_{1,2}|$ with $\xi_{1}>\xi_{2}$ so that $\varrho_{\text{s}}^{(2)}(0)$ is closer to the steady state. Here, we control $\omega_{0}$ with fixed $\{\alpha,\gamma,\Omega\}$ to guarantee the two states are under the same environmental conditions. However, adjusting $\{\alpha,\gamma,\Omega\}$ with fixed $\omega_{0}$ yields the same physical result~\cite{SupplementalMaterial}. The evolution of $\varrho_{\text{s}}^{(1)}(t)$ is governed by the $\mathcal{\hat{L}}$ with $\omega_{0}=\Omega$, which ensures the occurrence of LEPs; while $\varrho_{\text{s}}^{(2)}(t)$ is determined by a $\mathcal{\hat{L}}$ without LEPs. As displayed in Fig.~\ref{fig:fig3}(c), an
intersection point is clearly observed, indicating a non-Markovian-LEP-induced QMPE. Moreover, in Fig.~\ref{fig:fig3}(d), we show that, if $\varrho_{\text{s}}^{(1)}(t)$ is also governed by a $\mathcal{\hat{L}}$ without LEPs, no QMPE occurs.

As comparisons, we also provide the Markovian results. Solving Eq.~(\ref{eq:eq11}), we find $\varrho_{\text{s}}^{\text{M}}(t)=|\xi P_{\text{M}}(t)\rangle\langle\xi P_{\text{M}}(t)|$ and $|| \varrho_{\text{s}}^{\text{M}}(t)-\varrho_{\text{s}}^{\text{eq}}||_{\text{TD}}=\sqrt{1-e^{-|\xi|^{2}e^{-\gamma_{\text{M}} t}}}$ where $P_{\text{M}}(t)=e^{-\gamma_{\text{M}} t/2-i\omega_{0}t}$. In distinct contrast to the non-Markovian result, no LEP is found, which once again proves the acceleration advantage facilitated by the additional spectral singularities of $\mathcal{\hat{L}}$ is a purely non-Markovian effect. Thus, within the common Born-Markovian treatment, no QMPE is observed [see Figs.~\ref{fig:fig3} (c) and \ref{fig:fig3}(d)].

\section{Discussion and conclusion}\label{sec:sec6}
Our theoretical predictions could be experimentally tested in the nuclear magnetic resonance platform~\cite{Chen2021,PhysRevLett.133.173602} or the circuit quantum electrodynamics architecture~\cite{jk6y-55xp}. As demonstrated in Refs.~\cite{Chen2021,PhysRevLett.133.173602}, the system-pseudomode Hamiltonian $\hat{H}_{\text{sp}}$ can be experimentally simulated by using the iodotriuroethylene molecule served as a four-qubit quantum simulator. The $^{13}\text{C}$ nucleus corresponds to the system (a qubit), while the other three $^{19}\text{F}$ nuclei are mapped to the truncated bosonic field~\cite{Chen2021,PhysRevLett.133.173602} of the pseudomode. Especially, as experimentally tested in Ref.~\cite{jk6y-55xp}, for certain excitation-restriction situations, the pseudomode can be modeled by a readout resonator in a superconducting circuit, where $\mathcal{\hat{L}}$ can be experimentally realized as a $9\times 9$ matrix~\cite{jk6y-55xp}.

In summary, by introducing auxiliary pseudomodes to incorporate the non-Markovian effect, the dissipative dynamics of an open quantum system can be equivalently described by a Lindbladian evolution with an extended Liouvillian superoperator. Using these non-Markovianity-induced additional LEPs, we propose an experimentally friendly scheme to observe the QMPE. Although only harmonic oscillator systems are displayed, our scheme can be generalized to other models, such as the discrete-variable system in Supplemental Material~\cite{SupplementalMaterial}, without difficulties. Compared with Ref.~\cite{PhysRevLett.127.060401}, we do not need any unitary rotations on the initial state, because our acceleration capacity is induced by the non-Markovian LEPs. Enriching our understanding of the non-Markovian effect during nonequilibrium dynamics, our findings may have potential applications in optimizing the performances of quantum energy devices.

\section{Acknowledgments}
This work is supported by the National Natural Science Foundation of China (Grants No. 12375015 and No. 12247101), the Fundamental Research Funds for the Central Universities (Grant No. lzujbky-2024-jdzx06), and the Natural Science Foundation of Gansu Province (Grants No. 22JR5RA389 and No. 25JRRA799) and the ``111 Center" under Grant No. B20063.

\section{Data availability}
The data that support the findings of this article are not publicly available upon publication because it is not technically feasible and/or the cost of preparing, depositing, and hosting the data would be prohibitive within the terms of this research project. The data are available from the authors upon reasonable request.

\bibliography{reference}

\clearpage
\onecolumngrid
\begin{center}
{\large \bf Supplemental Material for ``Quantum Mpemba Effect by Non-Markovian Exceptional Points"}\\
\vspace{0.3cm}
\end{center}

\begin{center}
{\large \bf 1. The general expression of reduced density matrix for a Gaussian open quantum system with linear couplings}\\
\vspace{0.3cm}
\end{center}

In our work, we concentrate on the reduced dissipative dynamics of a quantum system which is coupled to a bosonic bath via a linear (bilinear) interaction. Such a dynamics is quite general, because only a few constraints are placed on the characteristics of the environment and the form of coupling~\cite{PhysRevLett.116.120402,PhysRevLett.113.200403,PhysRevA.95.020101,PhysRevA.85.062323}. A large number of well-known open quantum models, such as the Caldeira-Leggett model (the quantum Brownian motion)~\cite{PhysRevD.45.2843}, the spin-boson model~\cite{RevModPhys.59.1}, the damped Jaynes-Cummings model~\cite{PhysRevLett.99.160502} and the pure dephasing model~\cite{10.1098/rspa.1996.0029}, fall in the class discussed here.

The considered Hamiltonian in the interaction picture reads
\begin{equation}
\hat{H}_{\text{int}}(t)=\sum_{\text{j}}\hat{S}_{\text{j}}(t)\hat{B}_{\text{j}}(t),
\end{equation}
where $\hat{S}_{\text{j}}(t)\equiv e^{it\hat{H}_{\text{s}}}\hat{S}_{\text{j}}e^{-it\hat{H}_{\text{s}}}$ and $\hat{B}_{\text{j}}(t)\equiv e^{it\hat{H}_{\text{b}}}\hat{B}_{\text{j}}e^{-it\hat{H}_{\text{b}}}$ are the system's dissipation operators and bosonic operators of the bath in the interaction picture, respectively. In the main text, we take $\hat{S}_{1}=\hat{S}_{2}^{\dagger}=\hat{S}$ and $\hat{B}_{1}=\hat{B}_{2}^{\dagger}=\hat{B}$ for the sake of simplification. The density matrix of the whole system $\rho_{\text{tot}}(t)$ evolves according to the von Neumann equation as
\begin{equation}
\dot{\rho}_{\text{tot}}(t)=-i\sum_{\text{j}}[\hat{S}_{\text{jL}}(t)\hat{B}_{\text{jL}}(t)-\hat{S}_{\text{jR}}(t)\hat{B}_{\text{jR}}(t)]\rho_{\text{tot}}(t),
\end{equation}
where the subscripts L and R denote the operators acting on $\rho_{\text{tot}}(t)$ from the left and the right, respectively, e.g., $\hat{S}_{\text{jL}}(t)\hat{B}_{\text{jR}}(t)\rho_{\text{tot}}(t)=\hat{S}_{\text{j}}(t)\rho_{\text{tot}}(t)\hat{B}_{\text{j}}(t)$~\cite{PhysRevLett.116.120402,PhysRevLett.113.200403,PhysRevA.95.020101,PhysRevA.85.062323}.
By partially tracing out the degrees of freedom of the bath, the reduced density matrix of the system reads
\begin{equation}
\begin{split}
\rho_{\text{s}}(t)=&\text{Tr}_{\text{b}}\bigg{[}\mathcal{\hat{T}}\exp\bigg{\{}-i\int_{0}^{t}d\tau\sum_{\text{j}}[\hat{S}_{\text{jL}}(\tau)\hat{B}_{\text{jL}}(\tau)-\hat{S}_{\text{jR}}(\tau)\hat{B}_{\text{jR}}(\tau)]\bigg{\}}\rho_{\text{tot}}(0)\bigg{]}\\
=&\text{Tr}_{\text{b}}[\mathcal{\hat{T}}e^{-i\hat{\chi}}\rho_{\text{b}}(0)]\rho_{\text{s}}(0),
\end{split}
\end{equation}
where we have assumed that $\rho_{\text{tot}}(0)$ is an uncorrelated initial state $\rho_{\text{tot}}(0)=\rho_{\text{s}}(0)\otimes\rho_{\text{b}}(0)$, $\mathcal{\hat{T}}$ is the time-ordering operator and $\hat{\chi}\equiv\int_{0}^{t}d\tau\sum_{\text{j}}[\hat{S}_{\text{jL}}(\tau)\hat{B}_{\text{jL}}(\tau)-\hat{S}_{\text{jR}}(\tau)\hat{B}_{\text{jR}}(\tau)]$. Thanks to the Gaussianity of the bath~\cite{PhysRevLett.116.120402,PhysRevA.85.062323}, namely the initial state is Gaussian state and this Gaussianity can be fully preserved during the time evolution as a consequence of the bilinear interaction, one can apply the following identity (Wick's theorem~\cite{DIOSI1993517,PhysRevA.104.052209})
\begin{equation}
\text{Tr}_{\text{b}}[\mathcal{\hat{T}}e^{-i\hat{\chi}}\rho_{\text{b}}(0)]=\mathcal{\hat{T}}\exp\bigg{\{}-\frac{1}{2}\text{Tr}_{\text{b}}[\mathcal{\hat{T}}\hat{\chi}^{2}\rho_{\text{b}}(0)]\bigg{\}}.
\end{equation}
Then, the general expression of $\rho_{\text{s}}(t)$ is given by
\begin{equation}
\begin{split}
\rho_{\text{s}}(t)=&\mathcal{\hat{T}}\exp\bigg{\{}\int_{0}^{t}d\tau_{1}\int_{0}^{t}d\tau_{2}\sum_{\text{j,j}'}\text{Tr}_{\text{b}}[\hat{B}_{\text{j}}(\tau_{1})\hat{B}_{\text{j}'}(\tau_{2})\rho_{\text{b}}(0)][\hat{S}_{\text{j}',\text{L}}(\tau_{2})\hat{S}_{\text{j,R}}(\tau_{1})\\
&-\Theta(\tau_{1}-\tau_{2})\hat{S}_{\text{j,L}}(\tau_{1})\hat{S}_{\text{j}',\text{L}}(\tau_{2})-\Theta(\tau_{2}-\tau_{1})\hat{S}_{\text{j}',\text{R}}(\tau_{2})\hat{S}_{\text{j,R}}(\tau_{1})]\bigg{\}}\rho_{\text{s}}(0)\\
=&\mathcal{\hat{T}}\exp\bigg{\{}\int_{0}^{t}d\tau_{1}\int_{0}^{t}d\tau_{2}\sum_{\text{j,j}'}C_{\text{jj}'}(\tau_{1}-\tau_{2})[\hat{S}_{\text{j}',\text{L}}(\tau_{2})\hat{S}_{\text{j,R}}(\tau_{1})\\
&-\Theta(\tau_{1}-\tau_{2})\hat{S}_{\text{j,L}}(\tau_{1})\hat{S}_{\text{j}',\text{L}}(\tau_{2})-\Theta(\tau_{2}-\tau_{1})\hat{S}_{\text{j}',\text{R}}(\tau_{2})\hat{S}_{\text{j,R}}(\tau_{1})]\bigg{\}}\rho_{\text{s}}(0),
\end{split}
\end{equation}
where $\Theta(\tau_{1},\tau_{2})$ is the step function. The above expression shows that $\rho_{\text{s}}(t)$ depends only on the two-time correlation functions $C_{\text{jj}'}(t)$. The effects of multi-time bath correlation functions are converted into the combinations of two-time correlation functions due to the Wick's theorem, which guarantees that the Gaussian non-Markovian open system dynamics is tractable. For the non-Gaussian situation, such as the spin-bath models, the contributions of multi-time bath correlation functions should be taken into account~\cite{10.1063/1.5018725,PhysRevA.98.032116}. It is still an unknown question whether the employed method can be feasible in the non-Gaussian bath models. This may serve as an interesting direction for future research.

For the situation considered in the main text, namely, $\hat{S}_{1}=\hat{S}_{2}^{\dagger}=\hat{S}$, $\hat{B}_{1}=\hat{B}_{2}^{\dagger}=\hat{B}=\sum_{k}g_{k}\hat{b}_{k}^{\dagger}$ and the bath is initially prepared in the Fock vacuum state $\rho_{\text{b}}=\bigotimes_{k}|0_{k}\rangle\langle 0_{k}|$ (the zero-temperature bath case), we have
\begin{equation}
C_{11}(t)=C_{22}(t)=C_{12}(t)=0;~~~C_{21}(t)=\int_{0}^{\infty}d\omega J(\omega)e^{-i\omega t}.
\end{equation}
This result means, as long as the two-time correlation function $C_{21}(t)=C(t)$ or the spectral density $J(\omega)$ is known, the reduced density of the system shall be determined accordingly. Even though only the two-time bath correlation function $C(t)$ is involved in the expression of $\rho_{\text{s}}(t)$, due to the time-non-local convolution terms, which collect the non-Markovian effect, $\rho_{\text{s}}(t)$ still lacks a very simple expression except for only for very few models. Traditionally, certain approximations, such as the Born-Markovian and the secular approximations, are widely utilized. These approximations may provide a simple mathematical form, but they might miss some important physical features. In the main text, to overcome this difficulty of handling these time-non-local non-Markovian equation of motion, we employ the pseudomode master equation approach. The difficulty of solving non-Markovian dynamics is equivalently transformed into the challenge of dealing with an increased number of degrees of freedom in a time-local Markovian dynamics. From this perspective, the difficulty has been transformed, but not eliminated. However, the form of the Lindblad dynamics is fully preserved, which allows us to apply of traditional spectral analysis techniques.

\begin{center}
{\large \bf 2. The long-time behaviour of the trace distance}\\
\vspace{0.3cm}
\end{center}

In this section, we provide the details of derivation of Eq. (7) in the main text. Here, we focus on the case of the trace distance, but our formula can be straightforwardly generalized to other definitions of distance, such as the Hilbert-Schmidt distance, without difficulties. Using Eq. (6) in the main text and partially tracing out the degrees of freedom of the pseudomodes, one sees
\begin{equation}\label{eq:eqs1}
\varrho_{\text{s}}(t)=\sum_{\ell=0}^{M^{2}-1}\text{Tr}[l_{\ell}\rho_{\text{sp}}(0)]e^{\lambda_{\ell}t}r_{\ell}^{\text{s}},
\end{equation}
where $r_{\ell}^{\text{s}}=\text{Tr}_{\text{p}}(r_{\ell})$ is the reduced eigenmatrices~\cite{Lin2025}. Thus, the trace distance between $\varrho_{\text{s}}(t)$ and $\varrho_{\text{s}}^{\text{eq}}$ is then given by
\begin{equation}
\begin{split}
D(t)=&\|\varrho_{\text{s}}(t)-\varrho_{\text{s}}^{\text{eq}}\|_{\text{TD}}\\
=&\frac{1}{2}\text{Tr}\sqrt{\sum_{\ell,\ell'=1}^{M^{2}-1}\text{Tr}[l_{\ell}\rho_{\text{sp}}(0)]\text{Tr}[l_{\ell'}\rho_{\text{sp}}(0)]^{*}e^{(\lambda_{\ell}+\lambda_{\ell'}^{*})t}r_{\ell}^{\text{s}}(r_{\ell'}^{\text{s}})^{\dagger}}.
\end{split}
\end{equation}
If the Liouvillian superoperator has no spectral singularities (LEPs), one finds the long-time behaviour of $D(t)$ is determined by
\begin{equation}\label{eq:eqs9}
D(t)\propto\sqrt{e^{2\text{Re}(\lambda_{1})t}}=e^{-\Delta t},
\end{equation}
where we have arranged the eigenvalues in an ascending order as $0=\lambda_{0}<|\text{Re}(\lambda_{1})|\leq |\text{Re}(\lambda_{2})|\dots\leq|\text{Re}(\lambda_{M^{2}-1})|$. If the Liouvillian superoperator has one $\nu$-order LEP where $\nu<+\infty$, $\nu$ different eigenvalues and the corresponding eigenmatrices $\{\lambda_{\ell},r_{\ell}\}_{\ell\in\bar{\ell}}$ coalesce into $\{\bar{\lambda},\bar{r}\}$. In this case, the corresponding $\nu$-dimensional eigensubspace for the Liouvillian superoperator cannot be diagonalized. Fortunately, this problem can be solved by introducing the so-called generalized eigenmatrices~\cite{Lin2025,PRXQuantum.2.040346,Horn2012} $\{\bar{r}^{\text{s}}_{\nu_{1}}=\text{Tr}_{\text{p}}(\bar{r}_{\nu_{1}})\}$, which can be determined by solving a Jordan chain as $(\mathcal{\hat{L}}-\bar{\lambda})\bar{r}_{\nu_{1}}=\bar{r}_{\nu_{1}-1}$ with $\nu_{1}=1,2,...,\nu-1$ and $(\mathcal{\hat{L}}-\bar{\lambda})\bar{r}_{0}=0$. Then, Eq.~(\ref{eq:eqs1}) becomes~\cite{Lin2025,PRXQuantum.2.040346}
\begin{equation}
\varrho_{\text{s}}(t)\propto\sum_{\ell\notin \bar{\ell}}e^{\lambda_{\ell}t}r_{\ell}^{\text{s}}+e^{\bar{\lambda}t}\sum_{\nu_{1}=0}^{\nu-1}\sum_{\nu_{2}=0}^{\nu_{1}}\frac{t^{\nu_{2}}}{\nu_{2}!}\bar{r}^{\text{s}}_{\nu_{2}}.
\end{equation}
From the above expression, one finds the long-time behaviour of $D(t)$ is determined by
\begin{equation}\label{eq:eqs11}
D(t)\propto t^{\nu-1}e^{-\Delta t}\propto e^{-\Delta t}.
\end{equation}
Combining Eq.~(\ref{eq:eqs9}) and Eq.~(\ref{eq:eqs11}), one reproduces Eq. (7) in the main text, which is also numerically testified by the illustrative examples displayed in Fig. 2 and Fig. 3 of the main text.

\begin{center}
{\large \bf 3. Eigenvalues of $\mathcal{\hat{L}}$ and the general expression of $\rho_{\text{sp}}(t)$}\\
\vspace{0.3cm}
\end{center}

\subsection{3.1 Eigenvalues of $\mathcal{\hat{L}}$}

In this section, we provide the details of deriving the analytical expressions of eigenvalues. First, we rewrite the system-pseudomode Hamiltonian $\hat{H}_{\text{sp}}$ as
\begin{equation}
\begin{split}
\hat{H}_{\text{sp}}=&\pmb{\hat{a}}^{\dagger}\pmb{\Omega}\pmb{\hat{a}}\\
=&(\hat{a}^{\dagger}_{0},\hat{a}^{\dagger}_{1},\hat{a}^{\dagger}_{2},\cdots,\hat{a}^{\dagger}_{N})\left(
                                                                                                                   \begin{array}{ccccc}
                                                                                                                     \omega_{0} & \alpha_{1} & \alpha_{2} & \cdots & \alpha_{N} \\
                                                                                                                     \alpha_{1} & \Omega_{1} & 0 & \cdots & 0 \\
                                                                                                                     \alpha_{2} & 0 & \Omega_{2} & \cdots & 0 \\
                                                                                                                     \vdots & \vdots & \vdots & \ddots & \vdots \\
                                                                                                                     \alpha_{N} & 0 & 0 & \cdots & \Omega_{N} \\
                                                                                                                   \end{array}
                                                                                                                 \right)\left(
                                                                                                                          \begin{array}{c}
                                                                                                                            \hat{a}_{0} \\
                                                                                                                            \hat{a}_{1} \\
                                                                                                                            \hat{a}_{2} \\
                                                                                                                            \vdots \\
                                                                                                                            \hat{a}_{N} \\
                                                                                                                          \end{array}
                                                                                                                        \right),
\end{split}
\end{equation}
where $\pmb{\Omega}=\pmb{\Omega}^{\dag}$ is a $(N+1)\times(N+1)$ matrix. With the help of $\pmb{\Omega}$, the system-pseudomode master equation can be accordingly reexpressed as
\begin{equation}
\begin{split}
\dot{\rho}_{\text{sp}}(t)=&\mathcal{\hat{L}}\rho_{\text{sp}}(t)\\
=&-i\Bigg{[}\sum_{\imath,\jmath=0}^{N}\Omega_{\imath\jmath}\hat{a}^{\dagger}_{\imath}\hat{a}_{\jmath},\rho_{\text{sp}}(t)\Bigg{]}+\sum_{\imath,\jmath=0}^{N}\frac{1}{2}\gamma_{\imath\jmath}\left[2\hat{a}_{\jmath}\rho_{\text{sp}}(t)\hat{a}^{\dagger}_{\imath}-\rho_{\text{sp}}(t)\hat{a}^{\dagger}_{\imath}\hat{a}_{\jmath}-\hat{a}^{\dagger}_{\imath}\hat{a}_{\jmath}\rho_{\text{sp}}(t)\right],
\end{split}
\end{equation}
where $\Omega_{\imath\jmath}$ and $\gamma_{\imath\jmath}$ denote the matrix elements of $\pmb{\Omega}$ and $\pmb{\gamma}=\text{Diag}\{0,\gamma_{1},\gamma_{2},\dots,\gamma_{N}\}$, respectively.

To diagonalize $\mathcal{\hat{L}}$, we introduce the following superoperators~\cite{2lgr-34qp}
\begin{align}\label{sp}
\mathcal{\hat{N}}^{-}_{\imath\jmath}=\overleftarrow{\hat{a}^{\dagger}_{\imath}\hat{a}_{\jmath}}-\overrightarrow{\hat{a}^{\dagger}_{\imath}\hat{a}_{\jmath}},~~~\mathcal{\hat{K}}^{0}_{\imath\jmath}=\frac{1}{2}\left(\overleftarrow{\hat{a}^{\dagger}_{\imath}\hat{a}_{\jmath}}+\overrightarrow{\hat{a}_{\imath}^{\dagger}\hat{a}_{\jmath}}\right),~~~\mathcal{\hat{K}}^{-}_{\imath\jmath}=\overleftarrow{\hat{a}_{\jmath}}\overrightarrow{\hat{a}^{\dagger}_{\imath}},
\end{align}
where $\overrightarrow{\hat{x}}\hat{y}\equiv \hat{y}\hat{x}$ and $\overleftarrow{\hat{x}}\hat{y}\equiv \hat{x}\hat{y}$ are right-hand and left-hand action superoperators, respectively. Using them, the enlarged Liouvillian superoperator $\mathcal{\hat{L}}$ can be then reexpressed as
\begin{equation}\label{L}
\begin{split}
\mathcal{\hat{L}}=&\sum_{\imath,\jmath=0}^{N}\bigg{(}-i\Omega_{\imath\jmath}\mathcal{\hat{N}}^{-}_{\imath\jmath}+\gamma_{\imath\jmath}\mathcal{\hat{K}}^{-}_{\imath\jmath}-\gamma_{\imath\jmath}\mathcal{\hat{K}}^{0}_{\imath\jmath}\bigg{)}\\
=&\mathcal{\hat{N}}^{-}_{-i\pmb{\Omega}}+\mathcal{\hat{K}}^{-}_{\pmb{\gamma}}-\mathcal{\hat{K}}^{0}_{\pmb{\gamma}}.
\end{split}
\end{equation}
Here, we have used the notation
\begin{equation}
\mathcal{\hat{P}}_{\pmb{A}}\equiv\sum_{\imath,\jmath=0}^{N}A_{\imath\jmath}\mathcal{\hat{P}}_{\imath\jmath},
\end{equation}
where $\hat{\mathcal{P}}$ is an arbitrary superoperator and $A_{\imath\jmath}$ are the elements of complex-valued matrix $\pmb{A}$. As demonstrated in Ref.~\cite{2lgr-34qp}, using the structure of $\text{su}(1,1)$ Lie algebra of $\{\mathcal{\hat{N}}^{-},\mathcal{\hat{K}}^{-},\mathcal{\hat{K}}^{0}\}$, the superoperator $\mathcal{\hat{K}}_{\gamma}^{0}$ in $\mathcal{\hat{L}}$ can be eliminated via
\begin{equation}\label{eq:eqs6}
\mathcal{\hat{L}}_{\text{d}}=\exp(\mathcal{\hat{K}}_{\pmb{I}_{N}}^{-})\mathcal{\hat{L}}\exp(-\mathcal{\hat{K}}_{\pmb{I}_{N}}^{-})=\mathcal{\hat{N}}_{-i\pmb{\Omega}}^{-}-\mathcal{\hat{K}}_{\pmb{\gamma}}^{0},
\end{equation}
where $\pmb{I}_{N}$ is a $(N+1)$-dimensional identity operator. This result means $\mathcal{\hat{L}}_{\text{d}}$ can be expressed as
\begin{equation}
\mathcal{\hat{L}}_{\text{d}}=\overleftarrow{\hat{L}_{\text{eff}}}+\overrightarrow{\hat{L}^{\dagger}_{\text{eff}}},
\end{equation}
where $\hat{L}_{\text{eff}}=\pmb{\hat{a}}^{\dagger}\pmb{L}\pmb{\hat{a}}$. Here, $\pmb{L}$ is a $(N+1)\times (N+1)$ matrix which can be expressed as
\begin{equation}
\pmb{L}=\left(
          \begin{array}{ccccc}
            -i\omega_{0} & -i\alpha_{1} & -i\alpha_{2} & \cdots & -i\alpha_{N} \\
            -i\alpha_{1} & -i\Omega_{1}-\frac{1}{2}\gamma_{1} & 0 & \cdots & 0 \\
            -i\alpha_{2} & 0 & -i\Omega_{1}-\frac{1}{2}\gamma_{1} & \ldots & 0 \\
            \vdots & \vdots & \vdots & \ddots & \vdots \\
            -i\alpha_{N} & 0 & 0 & \cdots & -i\Omega_{N}-\frac{1}{2}\gamma_{N} \\
          \end{array}
        \right).
\end{equation}
Then, the original Liouvillian spectrum problem $\mathcal{\hat{L}}r=\lambda r$ reduces to the counterpart problem of $\mathcal{\hat{L}}_{\text{d}}r_{\text{d}}=\lambda r_{\text{d}}$ with $r_{\text{d}}=\exp(\mathcal{\hat{K}}_{\pmb{I}_{N}}^{-})r$.

The matrix $\pmb{L}$ can be diagonalized via a unitary matrix $\pmb{S}$ as
\begin{equation}
\boldsymbol{L}_{\text{d}}=\boldsymbol{S}^{\dagger}\boldsymbol{L}\boldsymbol{S}=\text{Diag}\{\tilde{\lambda}_{0},\tilde{\lambda}_{1},\dots,\tilde{\lambda}_{N}\},
\end{equation}
where the eigenfrequencies $\tilde{\lambda}_{\imath}$ are determined by the following determinant equation
\begin{equation}
\det(\boldsymbol{L}-\lambda\boldsymbol{I}_{N})=0.
\end{equation}
Thanks to the fact that $\pmb{L}-\lambda\boldsymbol{I}_{N}$ is a block matrix, i.e.,
\begin{equation}
\pmb{L}-\lambda\boldsymbol{I}_{N}=\left(
                                    \begin{array}{cc}
                                      \pmb{A}_{1\times 1} & \pmb{B}_{1\times N} \\
                                      \pmb{C}_{N\times 1} & \pmb{D}_{N\times N} \\
                                    \end{array}
                                  \right)
\end{equation}
we have
\begin{equation}
\begin{split}
\det(\pmb{L}-\lambda\boldsymbol{I}_{N})=&\det(\pmb{D}_{N\times N})\det(\pmb{A}_{1\times 1}-\pmb{B}_{1\times N}\pmb{D}_{N\times N}^{-1}\pmb{C}_{N\times 1})\\
=&\bigg{[}(-i\omega_{0}-\lambda)+\sum_{j=1}^{N}\frac{\alpha_{j}^{2}}{-i\Omega_{j}-\frac{1}{2}\gamma_{j}-\lambda}\bigg{]}\prod_{i=1}^{N}\bigg{(}-i\Omega_{j}-\frac{\gamma_{j}}{2}-\lambda\bigg{)}.
\end{split}
\end{equation}
The above determinant equation is equivalent to finding the roots of the characteristic polynomial
\begin{equation}
Q(\lambda)=(i\omega_{0}+\lambda)\prod_{i=1}^{N}\left(i\Omega_{i}+\frac{\gamma_{i}}{2}+\lambda\right)+\sum_{i=1}^{N}\alpha_{i}^{2}\prod_{j\neq i,j=1}^{N}\left(i\Omega_{j}+\frac{\gamma_{j}}{2}+\lambda\right),
\end{equation}
which reproduces Eq.~(9) in the main text.

Defining a set of new bosonic operators through the Bogoliubov transformation $\boldsymbol{\hat{b}}=\boldsymbol{S}^{\dagger}\boldsymbol{\hat{a}}$, we obtain
\begin{equation}\label{Ld}
\hat{L}_{\text{eff},\text{d}}=\sum_{\imath=0}^{N}\tilde{\lambda}_{\imath}\hat{b}_{\imath}^{\dagger}\hat{b}_{\imath}.
\end{equation}
Thus, one finds $\hat{L}_{\text{eff},\text{d}}|\pmb{m}\rangle=\pmb{m}|\pmb{m}\rangle$ and $\langle \pmb{n}|\hat{L}_{\text{eff},\text{d}}^{\dagger}=\langle\pmb{n}|\pmb{n}^{*}$, where
\begin{equation}
\pmb{m}=\sum_{\imath=0}^{N}m_{\imath}\tilde{\lambda}_{\imath},~~~\pmb{n}^{*}=\sum_{\imath=0}^{N}n_{\imath}\tilde{\lambda}_{\imath}^{*},
\end{equation}
are eigenvalues and
\begin{equation}
|\pmb{m}\rangle=\bigotimes_{\imath=0}^{N}|m_{\imath}\rangle=|m_{0}\rangle\otimes|m_{1}\rangle\otimes|m_{2}\rangle\cdots\otimes|m_{N}\rangle,~~~\langle \pmb{n}|=\bigotimes_{\imath=0}^{N}\langle n_{\imath}|=\langle n_{0}|\otimes\langle n_{1}|\otimes\langle n_{2}|\cdots\otimes\langle n_{N}|
\end{equation}
are multi-mode Fock states. With these results at hand, we have
\begin{equation}
\begin{split}
\mathcal{\hat{L}}_{\text{d}}|\pmb{m}\rangle\langle\pmb{n}|=&\hat{L}_{\text{eff},\text{d}}|\pmb{m}\rangle\langle\pmb{n}|+|\pmb{m}\rangle\langle\pmb{n}|\hat{L}_{\text{eff},\text{d}}^{\dagger}\\
=&\Bigg{[}\sum_{\imath=0}^{N}\text{Re}(\tilde{\lambda}_{\imath})(m_{\imath}+n_{\imath})+i\text{Im}(\tilde{\lambda}_{\imath})(m_{\imath}-n_{\imath})\Bigg{]}|\pmb{m}\rangle\langle\pmb{n}|,
\end{split}
\end{equation}
which means $\mathcal{\hat{L}}_{\text{d}} r_{\text{d}}=\lambda r_{\text{d}}$ with $r_{\text{d}}=|\pmb{m}\rangle\langle\pmb{n}|$ being the eigenmatrices and
\begin{equation}
\lambda_{\pmb{mn}}=\sum_{\imath=0}^{N}\text{Re}(\tilde{\lambda}_{\imath})(m_{\imath}+n_{\imath})+i\text{Im}(\tilde{\lambda}_{\imath})(m_{\imath}-n_{\imath})
\end{equation}
being the eigenvalues. Here, $\pmb{mn}\equiv\{m_{0},n_{0},m_{1},n_{1},...,m_{N},n_{N}\}$ is a multi-parameter index. The above expression reproduces the result in the main text.

\subsection{3.2 General expression of $\rho_{\text{sp}}(t)$}

Next, we derive the explicit expression for the density matrix $\rho_{\text{sp}}(t)$. Using Eq.~(\ref{eq:eqs6}), we have
\begin{equation}
\rho_{\text{sp}}(t)=e^{\mathcal{\hat{L}}t}\rho_{\text{sp}}(0)=e^{-\mathcal{\hat{K}}^{-}_{\pmb{I}_{N}}}e^{\mathcal{\hat{L}}_{\text{d}}t}e^{\mathcal{\hat{K}}^{-}_{\pmb{I}_{N}}}\rho_{\text{sp}}(0).
\end{equation}
By inserting the completeness relation $\sum_{\pmb{\tilde{m}}}|\pmb{\tilde{m}}\rangle\langle\pmb{\tilde{m}}|=\pmb{I}_{N}$, where
\begin{equation}
|\pmb{\tilde{m}}\rangle=\bigotimes_{\imath=0}^{N}|\tilde{m}_{\imath}\rangle
\end{equation}
with $\hat{a}_{\imath}^{\dagger}\hat{a}_{\imath}|\tilde{m}_{\imath}\rangle=\tilde{m}_{\imath}|\tilde{m}_{\imath}\rangle$, one finds the density matrix can be expressed as
\begin{equation}\label{DM}
\rho_{\text{sp}}(t)=\sum_{\boldsymbol{\tilde{m}},\boldsymbol{\tilde{m}'},\boldsymbol{\tilde{n}},\boldsymbol{\tilde{n}'}} \rho_{\boldsymbol{\tilde{m}'}\boldsymbol{\tilde{n}'}} U_{\boldsymbol{\tilde{m}}'\boldsymbol{\tilde{m}}}(t)U^{*}_{\boldsymbol{\tilde{n}}\boldsymbol{\tilde{n}'}}(t)\sigma_{\boldsymbol{\tilde{m}}\boldsymbol{\tilde{n}}},
\end{equation}
where
\begin{equation}
\begin{split}
\rho_{\boldsymbol{\tilde{m}}\boldsymbol{\tilde{n}}}=&\prod_{\imath=0}^{N}\exp(-\mathcal{\hat{K}}_{\pmb{I}_{1}}^{-})|\tilde{m}_{\imath}\rangle\langle \tilde{n}_{\imath}|\\
=&\prod_{\imath=0}^{N}\sum_{k=0}^{\min(\tilde{m}_{\imath},\tilde{n}_{\imath})}\frac{(-1)^{k}}{k!}\hat{a}_{\imath}^{k}|\tilde{m}_{\imath}\rangle\langle \tilde{n}_{\imath}|(\hat{a}^{\dagger}_{\imath})^{k}\\
=&\prod_{\imath=0}^{N}\sum_{k=0}^{\min(\tilde{m}_{\imath},\tilde{n}_{\imath})} (-1)^{k}\sqrt{\binom{\tilde{m}_{\imath}}{k}\binom{\tilde{n}_{\imath}}{k}}\ket{\tilde{m}_{\imath}-k}\bra{\tilde{n}_{\imath}-k},
\end{split}
\end{equation}
\begin{equation}
\begin{split}
\quad\sigma_{\boldsymbol{\tilde{m}}\boldsymbol{\tilde{n}}}=&\prod_{\imath=0}^{N}\langle \tilde{m}_{\imath}|\exp(\mathcal{\hat{K}}_{\pmb{I}_{1}}^{-})\rho_{\text{sp}}(0)|\tilde{n}_{\imath}\rangle\\
=&\prod_{\imath=0}^{N}\sum_{l=0}^{\infty} \frac{1}{l!} \bra{\tilde{m}_{\imath}}\hat{a}_{\imath}^{l}\rho_{\text{sp}}(0)(\hat{a}_{\imath}^{\dagger})^{l}\ket{\tilde{n}_{\imath}},
\end{split}
\end{equation}
\begin{equation}\label{r3}
\begin{split}
U_{\boldsymbol{\tilde{m}}\boldsymbol{\tilde{m}'}}(t)=&\bra{\boldsymbol{\tilde{m}}}e^{\hat{L}_{\text{eff}}t}\ket{\boldsymbol{\tilde{m}'}}\\
=&\bra{\boldsymbol{\tilde{0}}}\prod_{\imath=0}^{N}\frac{\hat{a}^{\tilde{m}_{\imath}}_{\imath}}{\sqrt{\tilde{m}_{\imath}!}}\prod_{\imath'=0}^{N}\frac{1}{\sqrt{\tilde{m}'_{\imath'}!}}\left[\sum_{\imath''}P_{\imath''\imath'}(t)\hat{a}^{\dagger}_{\imath''}\right]^{\tilde{m}'_{\imath'}}\ket{\boldsymbol{\tilde{0}}},
\end{split}
\end{equation}
with $P_{\imath\imath'}(t)=(e^{\boldsymbol{L}t})_{\imath\imath'}$.

\begin{center}
{\large \bf 4. Exact non-Markovian results}\\
\vspace{0.3cm}
\end{center}

In the Exemplification section of the main text, we consider a dissipative quantum harmonic oscillator, which is coupled to a Lorentzian bosonic bath, as an illustrative example. At zero temperature, the bath correlation function reads
\begin{equation}
\begin{split}
C(t)=&\int_{0}^{\infty}d\omega J(\omega)e^{-i\omega t}\\
\simeq&\int_{-\infty}^{\infty}\frac{d\omega}{2\pi}\frac{\Gamma \Lambda^{2}}{(\omega-\Omega)^{2}+\Lambda^{2}}e^{-i\omega t}\\
=&\frac{\Gamma\Lambda}{2}e^{-(\Lambda+i\Omega)t},
\end{split}
\end{equation}
where we have extended the integration range of $\omega$ to negative frequencies. In this case, one finds $\gamma=2\Lambda$, $\alpha=\sqrt{\Gamma\Lambda/2}$ and the corresponding expression of $\hat{L}_{\text{eff}}$ is then given by
\begin{equation}
\hat{L}_{\text{eff}}=-i\omega_{0}\hat{a}_{0}^{\dagger}\hat{a}_{0}-\left(\frac{\gamma}{2}+i\Omega\right)\hat{a}_{1}^{\dagger}\hat{a}_{1}-i\alpha (\hat{a}_{1}^{\dagger}\hat{a}_{0}+\hat{a}_{1}\hat{a}_{0}^{\dagger}),
\end{equation}
from which one finds
\begin{equation}
\pmb{L}=\begin{pmatrix}
        -i\omega_{0} & -i\alpha\\
        -i\alpha & -i\Omega-\frac{1}{2}\gamma\\
    \end{pmatrix}.
\end{equation}
With the above expression at hand, we find
\begin{equation}
\begin{split}
\rho_{\boldsymbol{\tilde{m}'}\boldsymbol{\tilde{n}'}}=&\prod_{\imath=0}^{1}\sum_{k=0}^{\min(\tilde{m}'_{\imath},\tilde{n}'_{\imath})}(-1)^{k}\sqrt{\binom{\tilde{m}'_{\imath}}{k}\binom{\tilde{n}'_{\imath}}{k}}\ket{\tilde{m}'_{\imath}-k}\bra{\tilde{n}'_{\imath}-k}
\end{split}
\end{equation}
\begin{align}
\sigma_{\boldsymbol{\tilde{m}}\boldsymbol{\tilde{n}}}=\frac{\xi^{\tilde{m}_{0}}}{\sqrt{\tilde{m}_{0}!}}\frac{\xi^{\tilde{n}_{0}}}{\sqrt{\tilde{n}_{0}!}}\delta_{\tilde{m}_{1}0}\delta_{\tilde{m}_{1}\tilde{n}_{1}},
\end{align}
and
\begin{equation}
U_{\boldsymbol{\tilde{m}}\boldsymbol{\tilde{m}'}}(t)=\sqrt{\frac{\tilde{m}_{0}!}{\tilde{m}'_{0}!\tilde{m}'_{1}!}}P_{00}^{\tilde{m}'_{0}}(t)P_{01}^{\tilde{m}'_{1}}(t)\delta_{\tilde{m}_{0},\tilde{m}'_{0}+\tilde{m}'_{1}},~~U_{\boldsymbol{\tilde{n}}\boldsymbol{\tilde{n}'}}(t)=\sqrt{\frac{\tilde{n}_{0}!}{\tilde{n}'_{0}!\tilde{n}'_{1}!}}P_{00}^{\tilde{n}'_{0}}(t)P_{01}^{\tilde{n}'_{1}}(t)\delta_{\tilde{n}_{0},\tilde{n}'_{0}+\tilde{n}'_{1}},
\end{equation}
where
\begin{equation}
P_{00}(t)=e^{-i(\omega_{0}+\Omega)t/2-\gamma t/4}\left[\cosh\left(\kappa t\right)+\frac{\gamma+2i(\Omega-\omega_{0})}{4\kappa}\sinh\left(\kappa t\right)\right],
\end{equation}
and
\begin{equation}
P_{01}(t)=P_{10}(t)=-i\frac{\alpha \sinh\left(\kappa t\right)}{\kappa}e^{-i(\omega_{0}+\Omega)t/2-\gamma t/4},
\end{equation}
with $\kappa=\sqrt{[\gamma+2i(\Omega-\omega_{0})]^{2}/16-\alpha^{2}}$. Substituting the above expressions into Eq.~(\ref{DM}) yields the expression of $\rho_{\text{sp}}(t)$. Partially tracing out the degrees of freedom of the pseudomodes, we find
\begin{equation}\label{eq:eqs36}
    \begin{split}
        \varrho_{\text{s}}(t)&=\sum_{\tilde{m}_{0},\tilde{n}_{0}=0}^{\infty}\sum_{k=0}^{\text{min}(\tilde{m}_{0},\tilde{n}_{0})}[\xi P_{00}(t)]^{\tilde{m}_{0}}[\xi P_{00}^{*}(t)]^{\tilde{n}_{0}}\frac{(-1)^{k}}{k!}\sqrt{\frac{1}{(\tilde{m}_{0}-k)!(\tilde{n}_{0}-k)!}}\ket{\tilde{m}_{0}-k}\bra{\tilde{n}_{0}-k}\\
        &=\sum_{l=0}^{\infty}\frac{(-1)^{l}|\xi P_{00}(t)|^{2l}}{l!}\sum_{\tilde{m}_0,\tilde{n}_0=0}^{\infty}\frac{[\xi P_{00}(t)]^{\tilde{m}_0}}{\sqrt{\tilde{m}_{0}!}}\frac{[\xi P_{00}^{*}(t)]^{\tilde{n}_0}}{\sqrt{\tilde{n}_{0}!}}\ket{\tilde{m}_0}\bra{\tilde{n}_0}\\
        &=\sum_{\tilde{m}_{0}=0}^{\infty}e^{-|\xi P_{00}(t)|^{2}/2}\frac{[\xi P_{00}(t)]^{\tilde{m}_{0}}}{\sqrt{\tilde{m}_{0}!}}\ket{\tilde{m}_{0}}\sum_{\tilde{n}_{0}=0}^{\infty}e^{-|\xi P_{00}(t)|^{2}/2}\frac{[\xi P_{00}^{*}(t)]^{\tilde{n}_{0}}}{\sqrt{\tilde{n}_{0}!}}\bra{\tilde{n}_{0}}\\
        &=\ket{\xi P_{00}(t)}\bra{\xi P_{00}(t)}.
    \end{split}
\end{equation}
Using the abbreviation $P(t)$ for $P_{00}(t)$, the result in the main text can be recovered.

On the other hand, one finds
\begin{equation}
\hat{L}_{\text{eff},\text{d}}=\tilde{\lambda}_{0}\hat{b}_{0}^{\dagger}\hat{b}_{0}+\tilde{\lambda}_{1}\hat{b}_{1}^{\dagger}\hat{b}_{1},
\end{equation}
with
\begin{equation}
\tilde{\lambda}_{0,1}=-\frac{1}{4}\gamma\pm\kappa-\frac{i}{2}(\omega_{0}+\Omega).
\end{equation}
From the above expression, one immediately sees that an LEP occurs at $\gamma=4\alpha$ and $\omega_{0}=\Omega$.

\begin{center}
{\large \bf 5. Comparisons with the traditional Born-Markovian results}\\
\vspace{0.3cm}
\end{center}

\subsection{5.1 Reproduction of the standard Born-Markovian master equation}

To recover the usual Born-Markovian master equation, one first needs to find the relation between the pseudomode operator $\hat{a}_{i}$ and the bosonic mode of system $\hat{a}_{0}$. This can be realized by using the so-called adjoint master equation approach proposed in Ref.~\cite{breuer2002theory}. To be specific, with the help of the system-pseudomode master equation
\begin{equation}\label{POM}
\begin{split}
\dot{\rho}_{\text{sp}}(t)=&\mathcal{\hat{L}}\rho_{\text{sp}}(t)\\
=&-i[\hat{H}_{\text{sp}},\rho_{\text{sp}}(t)]+\sum_{i=1}^{N}\gamma_{i}\left[\hat{a}_{i}\rho_{\text{sp}}(t)\hat{a}_{i}^{\dagger}-\frac{1}{2}\hat{a}_{i}^{\dagger}\hat{a}_{i}\rho_{\text{sp}}(t)-\frac{1}{2}\rho_{\text{sp}}(t)\hat{a}_{i}^{\dagger}\hat{a}_{i}\right],
\end{split}
\end{equation}
the equation of motion for an arbitrary system operator $\hat{O}(t)$ in the Heisenberg picture is given by the following adjoint master equation~\cite{breuer2002theory}
\begin{equation}
\begin{split}
\frac{d}{dt}\hat{O}(t)=&\mathcal{\hat{L}}^{\dagger}\hat{O}(t)\\
=&i[\hat{H}_{\text{sp}},\hat{O}(t)]+\sum_{i=1}^{N}\gamma_{i}\left[\hat{a}_{i}^{\dagger}\hat{O}(t)\hat{a}_{i}-\frac{1}{2}\hat{a}_{i}^{\dagger}\hat{a}_{i}\hat{O}(t)-\frac{1}{2}\hat{O}(t)\hat{a}_{i}^{\dagger}\hat{a}_{i}\right].
\end{split}
\end{equation}
Taking $\hat{O}=\hat{a}_{i}$, one easily finds the equation of motion for $\hat{a}_{i}(t)$ is given by
\begin{equation}
\begin{split}
\frac{d}{dt}\hat{a}_{i}(t)=\bigg{(}-i\Omega_{i}-\frac{\gamma_{i}}{2}\bigg{)}\hat{a}_{i}(t)-i\alpha \hat{a}_{0}.
\end{split}
\end{equation}
Solving the above equation yields
\begin{equation}
  \hat{a}_{i}(t)=-\frac{2i\alpha_{i}\hat{a}_{0}}{\gamma_{i}+2i\Omega_{i}}\left(1-e^{-i\Omega_{i}t-\frac{\gamma_{i}}{2}t}\right)+e^{-i\Omega_{i}t-\frac{\gamma_{i}}{2}t}\hat{a}_{i}(0).
\end{equation}
In the Markovian limit $\gamma_{i}\to\infty$, the above expression can be approximated as
\begin{equation}\label{a1}
  \hat{a}_{i}\approx -\frac{2i\alpha_{i}}{\gamma_{i}}\hat{a}_{0}.
\end{equation}

On the other hand, by employing the Born approximation, $\rho_{\text{sp}}(t)$ can be expressed as
\begin{equation}\label{eq:eq35}
\rho_{\text{sp}}(t)\approx \varrho_{\text{s}}^{\text{M}}(t)\otimes \rho_{\text{pm}}^{\text{eq}}=\varrho_{\text{s}}^{\text{M}}(t)\otimes \rho_{\text{pm}}(0),
\end{equation}
where $\rho_{\text{pm}}^{\text{eq}}$ denotes the long-time equilibrium state of pseudomodes. Substituting Eq. (\ref{a1}) and Eq. (\ref{eq:eq35}) into Eq. (\ref{POM}) and tracing out the degrees of freedom of pseudomodes, we obtain
\begin{equation}
\dot{\varrho}_{\text{s}}^{\text{M}}(t)=-i[\hat{H}_{\text{s}}+\hat{H}_{\text{LS}},\varrho_{\text{s}}^{\text{M}}(t)]+\sum_{i}\frac{4\alpha_{i}^{2}}{\gamma_{i}}\mathcal{\hat{D}}_{\hat{a}_{0}}[\varrho_{\text{s}}^{\text{M}}(t)],
\end{equation}
where
\begin{equation}
    \hat{H}_{\text{LS}}=\sum_{i}\frac{4\alpha_{i}^{2}\Omega_{i}}{\gamma_{i}^{2}}\hat{a}_{0}^{\dagger}\hat{a}_{0},
\end{equation}
is the Lamb-shift Hamiltonian. This result is in good agreement with the standard Lindblad equation in Ref. \cite{breuer2002theory}
\begin{equation}
\dot{\varrho}^{\text{M}}_{\text{s}}(t)=-i[\hat{H}_{\text{s}}+\hat{H}_{\text{LS}},\varrho^{\text{M}}_{\text{s}}(t)]+\gamma_{\text{M}}\mathcal{\hat{D}}_{\hat{a}_{0}}[\varrho^{\text{M}}_{\text{s}}(t)],
\end{equation}
where
\begin{equation}
    \gamma_{\text{M}}=2\text{Re}\left[\int_{0}^{\infty}dt C(t)\right]\simeq\sum_{i}\left(\frac{4\alpha_{i}^{2}}{\gamma_{i}}\right).
\end{equation}

\subsection{5.2 Reproduction of the Born-Markovian Liouvillian spectrum}

Next, we prove that the eigenvalues of $\hat{\mathcal{L}}$ naturally reduce to the eigenvalues of $\hat{\mathcal{L}}_{\text{M}}$ in the Markovian limit. From the characteristic polynomial $Q(\lambda)=0$, we have
\begin{equation}
    -i\omega_{0}-\sum_{i=1}^{N}\frac{\alpha_{i}^{2}}{i\Omega_{i}+\frac{1}{2}\gamma_{i}+\lambda}=\lambda.
\end{equation}
In the Markov limit $\gamma_{i}\to\infty$, one sees
\begin{equation}
\sum_{i=1}^{N}\frac{\alpha_{i}^{2}}{i\Omega_{i}+\frac{1}{2}\gamma_{i}+\lambda}\simeq\sum_{i=1}^{N}\frac{2\alpha_{i}^{2}}{\gamma_{i}}\bigg{(}1-\frac{\lambda+i\Omega_{i}}{\frac{1}{2}\gamma_{i}}\bigg{)}+o\bigg{(}\frac{1}{\gamma_{i}^{3}}\bigg{)}.
\end{equation}
Up to the lowest-order term, we find the following approximate solution
\begin{equation}
    \tilde{\lambda}_{0}^{\text{M}} \simeq -i\omega_{0}-\sum_{i}\frac{2\alpha^{2}_{i}}{\gamma_{i}}=-i\omega_{0}-\frac{\gamma_{\text{M}}}{2}.
\end{equation}
Thus, the non-Markovian Liouvillian spectrum reduces to the Markovian Liouvillian spectrum:
\begin{equation}
    \lambda_{\pmb{mn}}^{\text{M}}=-\frac{\gamma_{\text{M}}}{2}(m_{0}+n_{0})-i\omega_{0}(m_{0}-n_{0}),
\end{equation}
for $m_{0},n_{0}\in \mathbb{N}$, which matches the result in Ref.~\cite{longhi2025mpemba}.

\subsection{5.3 Expression of the density matrix under the Born-Markovian approximation}

Neglecting the Lamb-shift term $\hat{H}_{\text{LS}}$, $\hat{L}_{\text{eff}}$ is then given by $\hat{L}_{\text{eff}}=-(i\omega_{0}+\frac{1}{2}\gamma_{\text{M}})\hat{a}^{\dagger}_{0}\hat{a}_{0}$ under the Born-Markovian approximation. For the initial coherent state $\varrho^{\text{M}}_{\text{s}}(0)=\ket{\xi}\bra{\xi}$, we find
\begin{equation}\label{eq:eqs53}
    \begin{split}
        \varrho_{\text{s}}^{\text{M}}(t)&=\sum_{\tilde{m},\tilde{n}=0}^{\infty}e^{-i\omega_{0}(\tilde{m}-\tilde{n})t-\frac{1}{2}(\tilde{m}+\tilde{n})\gamma_{\text{M}}t}\xi^{\tilde{n}}\xi^{\tilde{m}}\sum_{l=0}^{\infty}\frac{(-1)^{l}}{l!}\sqrt{\frac{1}{(\tilde{n}-l)!(\tilde{m}-l)!}}\ket{\tilde{n}-l}\bra{\tilde{m}-l}\\
        &=\sum_{\tilde{m}_{0},\tilde{n}_{0},l=0}^{\infty}\frac{(-1)^{l}}{l!}\frac{1}{\sqrt{\tilde{m}_{0}!}}\xi^{\tilde{m}_{0}+l}e^{-(i\omega_{0}+\frac{1}{2}\gamma_{\text{M}})(\tilde{m}_{0}+l)t}\ket{\tilde{m}_{0}}\bra{\tilde{n}_{0}}e^{-(-i\omega_{0}+\frac{1}{2}\gamma_{\text{M}})(\tilde{n}_{0}+l)t}\xi^{\tilde{n}_{0}+l}\frac{1}{\sqrt{\tilde{n}_{0}!}}\\
        &=\ket{\xi e^{-(i\omega_{0}+\frac{1}{2}\gamma_{\text{M}})t}}\bra{\xi e^{-(i\omega_{0}+\frac{1}{2}\gamma_{\text{M}})t}},
    \end{split}
\end{equation}
which recovers the result in the main text, and in good agreement with that of the coherent-state path-integral method~\cite{PhysRevA.82.012105,PhysRevE.90.022122}.

\subsection{5.4 Measure of non-Markovianity}

\begin{figure}
\includegraphics[angle=0,width=0.3\textwidth]{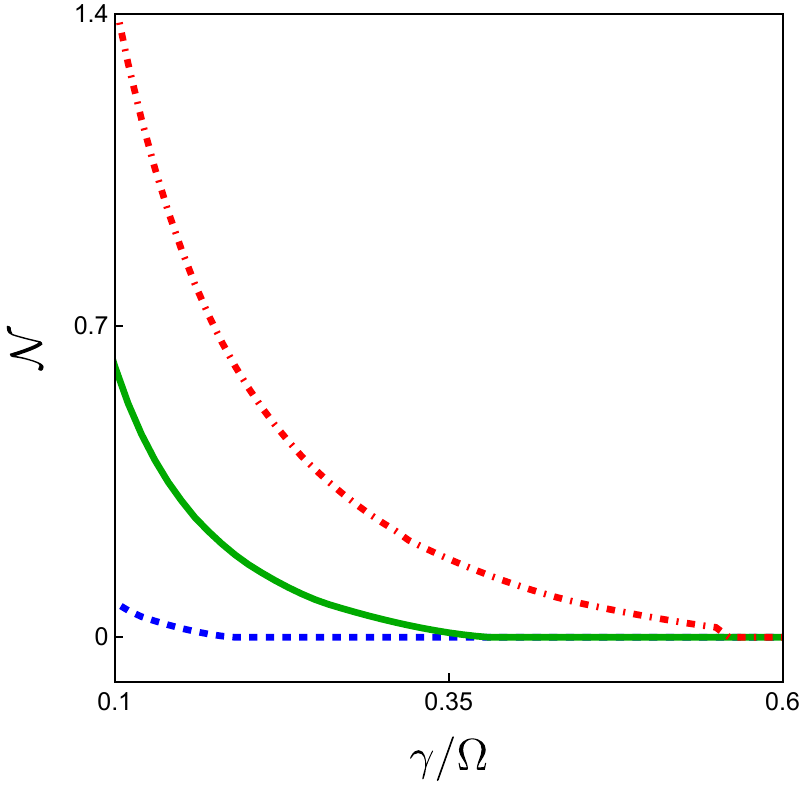}
\caption{The value of non-Markovianity is displayed as a function of $\gamma$ with different $\alpha$s: $\alpha=0.1\Omega$ (blue dashed line), $\alpha=0.2\Omega$ (green solid line) and $\alpha=0.3\Omega$ (red dot-dashed line). The initial coherent state parameters are $\xi_{1}=2$ and $\xi_{2}=1$, and other parameters are chosen as $\Omega=1~\text{cm}^{-1}$ and $\omega_{0}=1.2\Omega$.}\label{fig:figsm0}
\end{figure}

To demonstrate the above result (Eq.~(\ref{eq:eqs36})) from the pseudomode master equation approach truly capture the non-Markovian nature, we here use the Bures distance to quantify the degree of non-Markovianity~\cite{PhysRevA.102.022228}. Following Refs.~\cite{PhysRevA.102.022228,PhysRevLett.103.210401,PhysRevA.96.032125}, the emergence of the recoherence or the information backflow can be viewed as an evidence of the non-Markovianity, which can be reflected by the rate of change in the Bures distance. The degree of freedom of non-Markovianity can be measured by~\cite{PhysRevA.102.022228}
\begin{equation}\label{eq:eqsn}
\mathcal{N}=\int_{\sigma(t)>0}dt\sigma(t),
\end{equation}
where
\begin{equation}
\sigma(t)=\frac{d}{dt}D_{B}[\rho_{1}(t),\rho_{2}(t)].
\end{equation}
The time integration is extended over all time intervals $t\in[0,+\infty)$ in which $\sigma(t)$ is positive, and
\begin{equation}
D_{B}(\rho_{1},\rho_{2})=\sqrt{2-2\sqrt{\mathcal{F}(\rho_{1},\rho_{2})}},
\end{equation}
is the Bures distance with
\begin{equation}
\mathcal{F}(\rho_{1},\rho_{2})=\bigg{(}\text{Tr}\sqrt{\sqrt{\rho_{1}}\rho_{2}\sqrt{\rho_{1}}}\bigg{)}^{2}
\end{equation}
being the quantum fidelity. The definition of Eq.~(\ref{eq:eqsn}) depends critically on the choices of the initial-state pairs. To obtain a universal result, which is independent of the initial states, one further needs to maximize of $\mathcal{N}$ by running over all possible initial-state pairs $\rho_{1,2}(0)$~\cite{PhysRevLett.103.210401,PhysRevA.96.032125}. However, if one only needs to prove a specific dynamics is non-Markovian, such a maximization procedure is not required~\cite{PhysRevA.102.022228}. In the following, we omit this maximization and consider a fixed pair of initial coherent states.

The above definition of non-Markovianity is not suitable for a numerical simulation in the following two aspects: first, it is impossible to numerically simulate the dynamics of the subsystem from zero to infinity, which means we need a cutoff time $t_{c}$ for the time integration. Second, it is not very convenient to estimate whether or not $\sigma(t)$ is
positive at each given time. Thus, following Ref.~\cite{PhysRevA.96.032125}, we use an equivalent expression of the non-Markovianity as
\begin{equation}
\mathcal{N}=\frac{1}{2}\int_{0}^{t_{c}}dt[|\sigma(t)|+\sigma(t)].
\end{equation}

Let us consider two different initial coherent states, i.e., $\rho_{1}(0)=|\xi_{1}\rangle\langle\xi_{1}|$ and $\rho_{2}(0)=|\xi_{2}\rangle\langle\xi_{2}|$ with $\xi_{1,2}\in \mathbb{R}$. With the help of Eq.~(\ref{eq:eqs36}), one easily finds the corresponding expression of Bures distance is given by
\begin{equation}
D_{B}(\rho_{1},\rho_{2})=\sqrt{2-2e^{-\frac{1}{2}|(\xi_{1}-\xi_{2})P(t)|^{2}}}.
\end{equation}
With the above result at hand, the corresponding non-Markovianity $\mathcal{N}$ can be numerically computed accordingly. In Fig.~\ref{fig:figsm0}, we plot $\mathcal{N}$ versus $\gamma$. It is revealed that $\mathcal{N}>0$, which implies the decoherence process described by Eq.~(\ref{eq:eqs36}) is non-Markovian, in the small-$\gamma$ regimes. As $\gamma$ increases, $\mathcal{N}$ asymptotically approaches zero. This result is physically reasonable: the bath correlation function $\alpha(t)$ reduces to Dirac-$\delta$ function in the limit of $\gamma\rightarrow\infty$, which means the decoherence becomes Markovian. In fact, if one uses the Markovian approximation, Eq.~(\ref{eq:eqs36}) reduces to Eq.~(\ref{eq:eqs53}) and the corresponding expression of $\sigma(t)$ is given by
\begin{equation}
\sigma_{\text{M}}(t)=-\frac{\gamma_{\text{M}}|\xi_{1}-\xi_{2}|^{2}}{2\sqrt{2-2e^{-\frac{1}{2}|\xi_{1}-\xi_{2}|^{2}\gamma_{\text{M}}t}}}\exp\bigg{(}\frac{1}{2}|\xi_{1}-\xi_{2}|^{2}e^{-\gamma_{\text{M}}t}-e^{-\frac{1}{2}|\xi_{1}-\xi_{2}|^{2}\gamma_{\text{M}}t}-\gamma_{\text{M}}t\bigg{)},
\end{equation}
which is always negative. This expression means the non-Markovianity $\mathcal{N}=0$ permanently.

\subsection{5.5 Markovian LEPs versus non-Markovian LEPs}

In this section, we shall demonstrate that our proposed non-Markovian-LEP-induced QMPE scheme has certain advantages over that of the traditional strategy based on Markovian LEPs. The main advantage is that the condition of forming a LEP can be substantially relaxed due to the additional adjustability of pseudomodes.

\begin{figure}
\includegraphics[angle=0,width=0.975\textwidth]{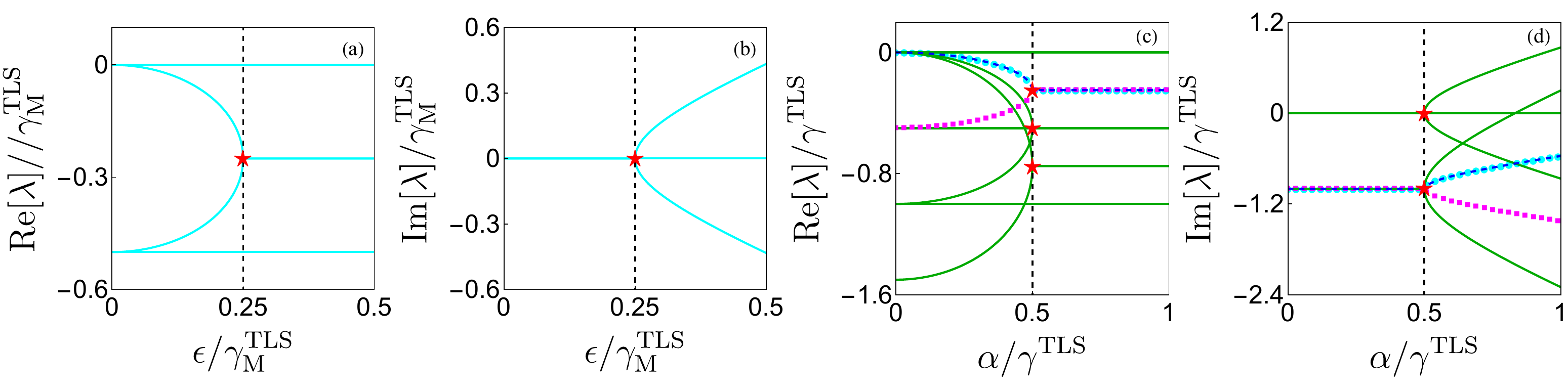}
\caption{Real (a) and imaginary (b) parts of the eigenvalues of $\mathcal{\hat{L}}_{\text{TLS}}^{\text{M}}$. As comparisons, real (c) and imaginary(b) parts of the eigenvalues of $\mathcal{\hat{L}}_{\text{TLS}}$ with $\Omega=\epsilon=1 \text{cm}^{-1}$. The LEPs are marked by the red five-pointed stars. The cyan circles and magenta rectangles in (c) and (d), respectively, present the eigenvalues of $\lambda_{1}$ and $\lambda_{2}$ from the analytical predictions of Eq.~(\ref{eq:eqs49}) in the low-dimensional subspace, while the green solid lines are numerical results from the full Liouvillian superoperator.}\label{fig:figsm1}
\end{figure}

Within the traditional Born-Markovian dynamics, the reduced dynamics of a dissipative two-level system, described by $\hat{H}_{\text{TLS}}=\frac{1}{2}\epsilon\hat{\sigma}_{z}$, is governed by the following standard Lindblad master equation
\begin{equation}
\begin{split}
\dot{\varrho}_{\text{TLS}}(t)=&\mathcal{\hat{L}}^{\text{M}}_{\text{TSL}}\varrho_{\text{TLS}}(t)\\
=&-i[\hat{H}_{\text{TLS}},\varrho_{\text{TLS}}(t)]+\frac{1}{4}\gamma^{\text{TLS}}_{\text{M}}\bigg{[}\hat{\sigma}_{x}\varrho_{\text{TLS}}(t)\hat{\sigma}_{x}-\frac{1}{2}\{\hat{\sigma}_{x}\hat{\sigma}_{x},\varrho_{\text{TLS}}(t)\}\bigg{]},
\end{split}
\end{equation}
where we have assumed the Lindblad dissipation operator is $\frac{1}{2}\hat{\sigma}_{x}$ and the decay rate reads $\gamma^{\text{TLS}}_{\text{M}}=2\pi J(\epsilon)$ with $J(\omega)$ being the spectral density. For the Lorentzian spectral density considered in the main text, we have $\gamma^{\text{TLS}}_{\text{M}}=\Gamma$ with $\Gamma$ being the coupling strength between the two-level system and the bath. The eigenvalues and the corresponding (unnormalized) eigenmatrices of $\mathcal{\hat{L}}^{\text{M}}_{\text{TSL}}$ are respectively given by
\begin{equation}
\lambda_{\ell}^{\text{M}}=\bigg{\{}0,-\frac{1}{4}\bigg{(}\gamma^{\text{TLS}}_{\text{M}}\pm\kappa^{\text{TLS}}_{\text{M}}\bigg{)},-\frac{1}{2}\gamma^{\text{TLS}}_{\text{M}}\bigg{\}},
\end{equation}
\begin{equation}
    r_{0}^{\text{M}}\propto \begin{pmatrix}
        1&0\\
        0&1
    \end{pmatrix},~~
    r_{1}^{\text{M}}\propto\begin{pmatrix}
        0&-4i\epsilon/\gamma^{\text{TLS}}_{\text{M}}-\kappa^{\text{TLS}}_{\text{M}}/\gamma^{\text{TLS}}_{\text{M}}\\
        1&0
    \end{pmatrix},~~
    r_{2}^{\text{M}}\propto\begin{pmatrix}
        0&-4i\epsilon/\gamma^{\text{TLS}}_{\text{M}}+\kappa^{\text{TLS}}_{\text{M}}/\gamma^{\text{TLS}}_{\text{M}}\\
        1&0
    \end{pmatrix},~~
    r_{3}^{\text{M}}\propto \begin{pmatrix}
        -1&0\\
        0&1
    \end{pmatrix},
\end{equation}
where $\kappa^{\text{TLS}}_{\text{M}}=\sqrt{(\gamma^{\text{TLS}}_{\text{M}})^{2}-16\epsilon^{2}}$. From the above expressions, one immediately sees a Markovian LEP, where $\lambda_{1}^{\text{M}}=\lambda_{2}^{\text{M}}$ and $r_{1}^{\text{M}}=r_{2}^{\text{M}}$, can be observed if the condition of $\gamma^{\text{TLS}}_{\text{M}}=4\epsilon$, or equivalently, $\Gamma=4\epsilon$ can be satisfied (see Fig.~\ref{fig:figsm1}).

On the other hand, using the pseudomode approach displayed in the main text, the equation of motion for the same two-level system becomes
\begin{equation}
\begin{split}
\dot{\rho}_{\text{TLS-P}}(t)=&\mathcal{\hat{L}}_{\text{TLS}}\rho_{\text{TLS-P}}(t)\\
=&-i[\hat{H}_{\text{TLS-P}},\rho_{\text{TLS-P}}(t)]+\gamma^{\text{TLS}}\bigg{[}\hat{a}\rho_{\text{TLS-P}}(t)\hat{a}^{\dagger}-\frac{1}{2}\{\hat{a}^{\dagger}\hat{a},\rho_{\text{TLS-P}}(t)\}\bigg{]},
\end{split}
\end{equation}
where
\begin{equation}\label{eq:eqhsp}
\hat{H}_{\text{TLS-P}}=\frac{1}{2}\epsilon\hat{\sigma}_{z}+\Omega\hat{a}^{\dagger}\hat{a}+\frac{1}{2}\alpha\hat{\sigma}_{x}(\hat{a}^{\dagger}+\hat{a})
\end{equation}
is the enlarged system-pseudomode Hamiltonian. Eq.~(\ref{eq:eqhsp}) is the well-known quantum Rabi model and can be experimentally simulated by trapped ions in the Lamb-Dicke limit~\cite{Zhang_2021,PhysRevResearch.5.043036}. For the Lorentzian spectral density, we have $\gamma^{\text{TLS}}=2\Lambda$, $\Omega=\epsilon$ and $\alpha=\sqrt{\Gamma\Lambda/2}$. One can easily find the spectrum of $\mathcal{\hat{L}}_{\text{TLS}}$ via a numerical diagonalization. However, to build a more clear physical picture, following Ref.~\cite{PhysRevResearch.5.043036}, we here reduce
$\hat{H}_{\text{TLS-P}}$ into a low-dimensional subspace of $\{|g0\rangle, |g1\rangle,|e0\rangle, |e,1\rangle\}$ with $|g,e\rangle$ and $|0,1\rangle$ being the eigenstates of $\hat{\sigma}_{z}$ and $\hat{a}^{\dagger}\hat{a}$, respectively. Such a treatment would be very helpful to obtain analytical results of eigenvalues and eigenmatrices. In the finite subspace, we find
\begin{equation}\label{eq:eqs49}
\lambda_{\ell}=\bigg{\{}0,-\frac{1}{4}(\gamma^{\text{TLS}}\pm\kappa^{\text{TLS}})\pm i\epsilon,-\frac{1}{2}(\gamma^{\text{TLS}}\pm\kappa^{\text{TLS}}),-\frac{3}{4}(\gamma^{\text{TLS}}\pm\kappa^{\text{TLS}})\pm i\epsilon,-\frac{1}{2}\gamma^{\text{TLS}}\pm i\epsilon,-\frac{1}{2}\gamma^{\text{TLS}},-\gamma^{\text{TLS}}\bigg{\}},
\end{equation}
\begin{equation}
    r_{0}= \begin{pmatrix}
        0&0&0&0\\
        0&0&0&0\\
        0&0&0&0\\
        0&0&0&1
    \end{pmatrix},~~
    r_{1}\propto \begin{pmatrix}
        0&0&0&0\\
        0&0&0&i(\gamma^{\text{TLS}}-\kappa^{\text{TLS}})/(2\alpha)\\
        0&0&0&1\\
        0&0&0&0
    \end{pmatrix},~~
    r_{2}\propto \begin{pmatrix}
        0&0&0&0\\
        0&0&0&i(\gamma^{\text{TLS}}+\kappa^{\text{TLS}})/(2\alpha)\\
        0&0&0&1\\
        0&0&0&0
    \end{pmatrix},
\end{equation}
\begin{equation}
    r_{3}\propto \begin{pmatrix}
        0&0&0&0\\
        0&0&0&0\\
        0&0&0&0\\
        0&-i(\gamma^{\text{TLS}}-\kappa^{\text{TLS}})/(2\alpha)&1&0
    \end{pmatrix},~~
    r_{4}\propto \begin{pmatrix}
        0&0&0&0\\
        0&0&0&0\\
        0&0&0&0\\
        0&-i(\gamma^{\text{TLS}}+\kappa^{\text{TLS}})/(2\alpha)&1&0
    \end{pmatrix},
\end{equation}
\begin{equation}
    r_{5}\propto \begin{pmatrix}
        0&0&0&0\\
        0&-1/2+\kappa^{\text{TLS}}/(2\gamma^{\text{TLS}})&-i\alpha/\gamma^{\text{TLS}}&0\\
        0&i\alpha/\gamma^{\text{TLS}}&-1/2-\kappa^{\text{TLS}}/(2\gamma^{\text{TLS}})&0\\
        0&0&0&1
    \end{pmatrix},~~
    r_{6}\propto \begin{pmatrix}
        0&0&0&0\\
        0&-1/2-\kappa^{\text{TLS}}/(2\gamma^{\text{TLS}})&-i\alpha/\gamma^{\text{TLS}}&0\\
        0&i\alpha/\gamma^{\text{TLS}}&-1/2+\kappa^{\text{TLS}}/(2\gamma^{\text{TLS}})&0\\
        0&0&0&1
    \end{pmatrix},
\end{equation}
\begin{equation}
    r_{7}\propto \begin{pmatrix}
        0&1&i(\gamma^{\text{TLS}}+\kappa^{\text{TLS}})/(2\alpha)&0\\
        0&0&0&-i(\gamma^{\text{TLS}}+\kappa^{\text{TLS}})/(2\alpha)\\
        0&0&0&1\\
        0&0&0&0
    \end{pmatrix},~~
    r_{8}\propto \begin{pmatrix}
        0&1&-i(\gamma^{\text{TLS}}+\kappa^{\text{TLS}})/(2\alpha)&0\\
        0&0&0&-i(\gamma^{\text{TLS}}-\kappa^{\text{TLS}})/(2\alpha)\\
        0&0&0&1\\
        0&0&0&0
    \end{pmatrix},
\end{equation}
\begin{equation}
    r_{9}\propto \begin{pmatrix}
        0&0&0&0\\
        1&0&0&0\\
        -i(\gamma^{\text{TLS}}+\kappa^{\text{TLS}})/(2\alpha)&0&0&0\\
        0&i(\gamma^{\text{TLS}}+\kappa^{\text{TLS}})/(2\alpha)&1&0
    \end{pmatrix},~~
    r_{10}\propto \begin{pmatrix}
        0&0&0&0\\
        1&0&0&0\\
        -i(\gamma^{\text{TLS}}-\kappa^{\text{TLS}})/(2\alpha)&0&0&0\\
        0&i(\gamma^{\text{TLS}}-\kappa^{\text{TLS}})/(2\alpha)&1&0
    \end{pmatrix},
\end{equation}
\begin{equation}
    r_{11}= \begin{pmatrix}
        0&0&0&1\\
        0&0&0&0\\
        0&0&0&0\\
        0&0&0&0
    \end{pmatrix},~~
    r_{12}= \begin{pmatrix}
        0&0&0&0\\
        0&0&0&0\\
        0&0&0&0\\
        1&0&0&0
    \end{pmatrix},~~
    r_{13}\propto \begin{pmatrix}
        0&0&0&0\\
        0&-1/2&-i\gamma^{\text{TLS}}/(2\alpha)&0\\
        0&0&-1/2&0\\
        0&0&0&1
    \end{pmatrix},
\end{equation}
\begin{equation}
     r_{14}\propto \begin{pmatrix}
        0&0&0&0\\
        0&0&1&0\\
        0&1&0&0\\
        0&0&0&0
    \end{pmatrix},~~
    r_{15}\propto \begin{pmatrix}
        1&0&0&0\\
        0&-1&0&0\\
        0&0&-1&0\\
        0&0&0&1
    \end{pmatrix},
\end{equation}
where $\kappa^{\text{TLS}}=\sqrt{(\gamma^{\text{TLS}})^{2}-4\alpha^{2}}$.

\begin{figure}
\includegraphics[angle=0,width=0.55\textwidth]{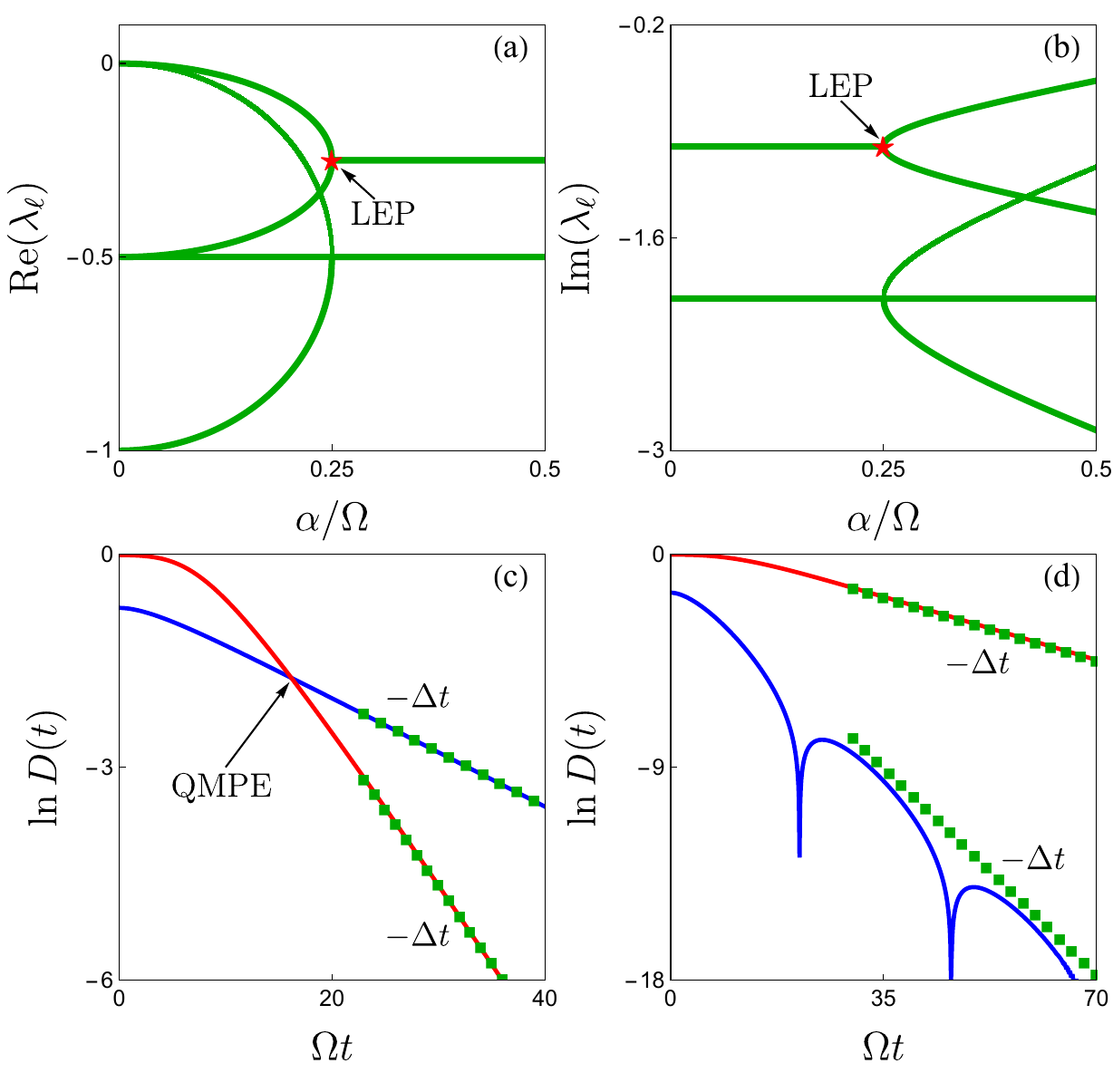}
\caption{Real (a) and imaginary (b) parts of the first five nonzero eigenvalues of $\mathcal{\hat{L}}$. (c) $\ln D(t)$ is plotted as a function of $\Omega t$ with different parameters: $\alpha=0.18$ and $\xi=0.5$ (blue solid line) and $\alpha=0.25$ and $\xi=2$ (red solid line). Subfigure (d) is the same as (c), but $\alpha=0.28$ and $\xi=0.2$ (blue solid line) and $\alpha=0.18$ and $\xi=2$ (red solid line). The red star corresponds to the first LEP where $\lambda_{1}=\lambda_{2}$. The green diamonds are analytical results of $\ln D(t)\propto-\Delta t$. Other parameters are chosen as $\gamma=\omega_{0}=\Omega=1~\text{cm}^{-1}$.}\label{fig:figsm2}
\end{figure}

From the above expressions, one sees a non-Markovian LEP, where $\lambda_{1}=\lambda_{2}$ and $r_{1}=r_{2}$, occurs at $\gamma^{\text{TLS}}=2\alpha$, or equivalently, $\Gamma=2\Lambda$. This analytical prediction matches very well with the numerical results calculated from the full size (see Fig.~\ref{fig:figsm1}). Due to the fact that the frequency of two-level system $\epsilon$ is much larger than the spectral width $\Lambda$, for example, in the recent experiment~\cite{jk6y-55xp}, $\epsilon\sim \text{GHz}$ while $\Lambda\sim \text{MHz}$, one concludes that the condition of finding a LEP can be greatly relaxed. In this sense, our proposed scheme can realize a QMPE via a much smaller system-bath coupling, which is more experimentally feasible.

\begin{center}
{\large \bf 6. Observation of QMPE by controlling the parameters of bath}\\
\vspace{0.3cm}
\end{center}

For the same model considered in Exemplification of the main text, one can also observe the QMPE by changing the parameters of bath $\{\alpha,\gamma,\Omega\}$ with fixed frequency $\omega_{0}=\Omega$. We here provide one figure (Fig. \ref{fig:figsm2}) to demonstrate this conclusion. We here display the eigenvalues of $\mathcal{\hat{L}}$ as a function of $\gamma$. Clearly, an LEP can be revealed if $\gamma=4\alpha$. Using this LEP-induced acceleration, the QMPE is observed (see Fig. \ref{fig:figsm2}(c)). Without the acceleration mechanism induced by LEPs, the QMPE would not occur (see Fig. \ref{fig:figsm2}(d)).

\begin{center}
{\large \bf 7. Application in quantum battery}\\
\vspace{0.3cm}
\end{center}

\begin{figure}
\includegraphics[angle=0,width=0.3\textwidth]{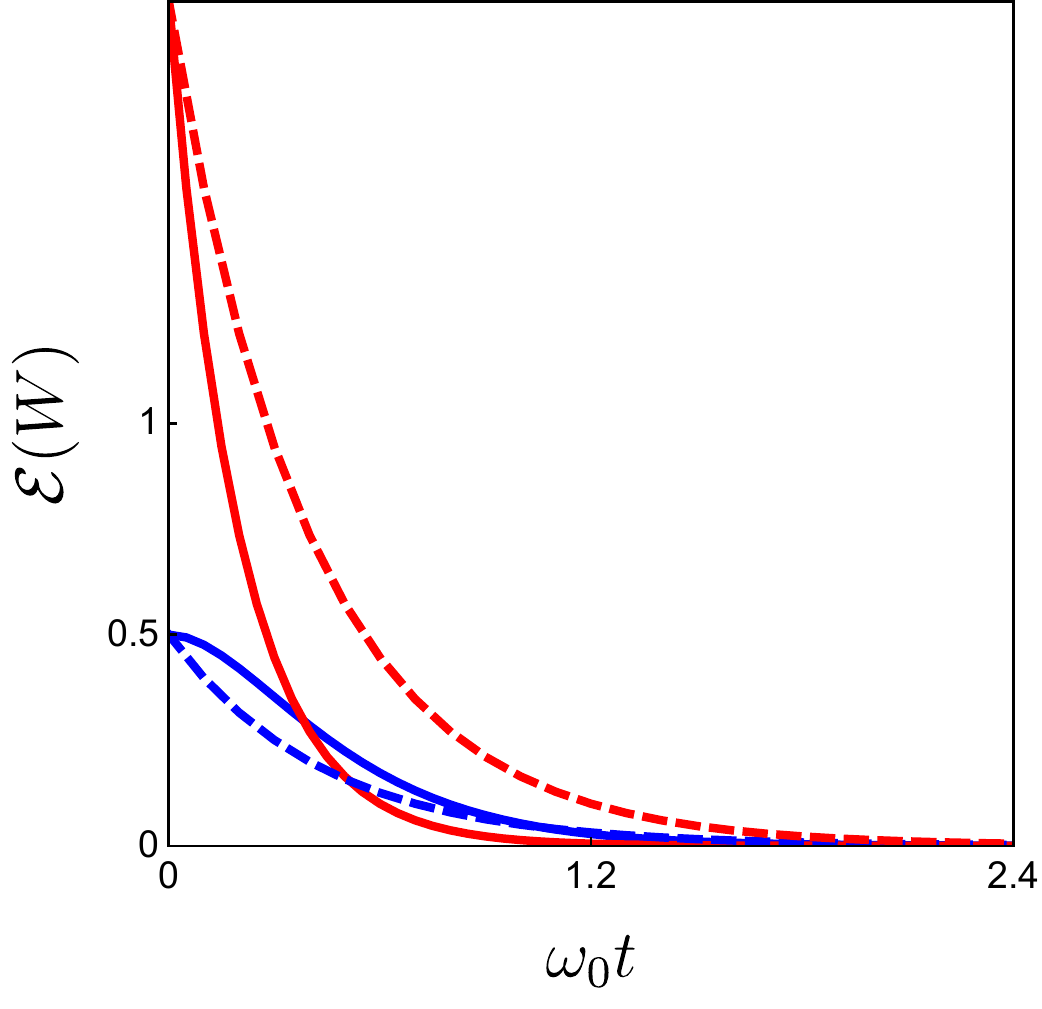}
\caption{The discharging process of quantum battery described by $\mathcal{E}(t)$ for both non-Markovian (solid lines) and Markovian (dashed lines) predictions. Parameters are the same with those of Fig.~\ref{fig:figsm2}.}\label{fig:figsm3}
\end{figure}

In this section, we apply our finding to speedup the discharging process of a quantum-harmonic-oscillator battery. The maximum of the extractable energy in quantum battery called ergotropy is defined as~\cite{AEAllahverdyan_2004,PhysRevE.102.042111}
\begin{equation}
\mathcal{E}(t)=\text{Tr}[\varrho_{\text{s}}(t)\hat{H}_{\text{s}}]-\text{Tr}(\pi_{\text{s}}\hat{H}_{\text{s}}),
\end{equation}
where $\pi_{\text{s}}$ is the thermal passive state of the system with inverse temperature $\beta_{\text{s}}$. The ergotropy is one of the most important indexes to evaluate the performance of quantum battery. When $\varrho_{\text{s}}(t)$ is a Gaussian state, it can be fully described by a Wigner function as
\begin{equation}
W(\pmb{\zeta})=\frac{1}{\pi\sqrt{|\pmb\sigma|}}\exp\bigg{[}-\frac{1}{2}(\pmb{\zeta}-\pmb{d})^{\text{T}}\pmb\sigma^{-1}(\pmb{\zeta}-\pmb{d})\bigg{]},
\end{equation}
where $\pmb{\zeta}=(\zeta,\zeta^{*})$, matrix elementals for the first and the second moment are, respectively, $d_{\text{i}}=\langle \hat{R}_{\text{i}}\rangle$ and $\sigma_{\text{ij}}=\frac{1}{2}\langle\{\hat{R}_{\text{i}},\hat{R}^{\dagger}_{\text{j}}\}\rangle-\langle \hat{R}_{\text{i}}\rangle\langle \hat{R}_{\text{j}}^{\dagger}\rangle$ with $\hat{\pmb{R}}=(\hat{a},\hat{a}^{\dagger})$. In this case, the ergotropy can be rewritten as~\cite{PhysRevLett.134.220402}
\begin{equation}
\mathcal{E}(t)=\frac{1}{2}\omega_{0}\bigg{(}\bar{n}_{\text{s}}+\frac{1}{2}\bigg{)}K[W(\pmb{\zeta})\parallel W_{\pi}],
\end{equation}
where $W_{\pi}$ is the Wigner function for the thermal passive state, and $K[W(\pmb{\zeta})\parallel W_{\pi}]$ is the relative Wigner entropy. As discussed in the Exemplification, we assume the temperature of the bath is zero and the initial state is a coherent state, we find
\begin{equation}
\mathcal{E}(t)=\frac{1}{2}\omega_{0}|\xi P(t)|^{2},
\end{equation}
which describes the discharging process of the quantum battery.

Next, we choose two different initial coherent states $\varrho_{\text{s}}^{(1,2)}(0)=|\xi_{1,2}\rangle\langle\xi_{1,2}|$ with $\xi_{1}>\xi_{2}$. Clearly, $\varrho_{\text{s}}^{(2)}(0)$ closer to the steady state. The first discharging process is governed by $\mathcal{\hat{L}}$ with a LEP. The second discharging process is determined by $\mathcal{\hat{L}}$ without LEPs. As displayed in Fig.~\ref{fig:figsm3}, the first discharging rate is much larger than the second one. Such an acceleration advantage disappears if one uses the Born-Markovian approximation. These results imply that the non-Markovian-LEP-induced QMPE can be employed to accelerate the discharging of a quantum battery.

\end{document}